\documentclass[fleqn,usenatbib]{mnras}

\usepackage{newtxtext,newtxmath}
\usepackage{rotating}
\usepackage{graphicx}

\usepackage[T1]{fontenc}

\DeclareRobustCommand{\VAN}[3]{#2}
\let\VANthebibliography\thebibliography
\def\thebibliography{\DeclareRobustCommand{\VAN}[3]{##3}\VANthebibliography}

\newcommand{\dx}{\Delta_\mathrm{x}}
\newcommand{\dy}{\Delta_\mathrm{y}}
\newcommand{\dbi}{\Delta_\mathrm{BI}}
\newcommand{\dcubi}{\Delta_\mathrm{C\,UBI}}

\newcommand{\mua}{\mu_\mathrm{\alpha}^*}
\newcommand{\mud}{\mu_\mathrm{\delta}}
\newcommand{\mur}{\mu_\mathrm{R}}
\newcommand{\mut}{\mu_\mathrm{T}}
\newcommand{\mux}{\mu_\mathrm{x}}
\newcommand{\muy}{\mu_\mathrm{y}}
\newcommand{\ani}{\sigma_\mathrm{T}/\sigma_\mathrm{R} -1}
\newcommand{\st}{\sigma_\mathrm{T}}
\newcommand{\sr}{\sigma_\mathrm{R}}

\newcommand{\rh}{R_\mathrm{h}}
\newcommand{\rc}{R_\mathrm{c}}
\newcommand{\rt}{R_\mathrm{t}}
\newcommand{\rj}{R_\mathrm{J}}

\newcommand{\vesc}{v_\mathrm{esc}}
\newcommand{\cubi}{C_\mathrm{UBI}}

\newcommand{\typeii}{Type {\rm II} }

\usepackage{graphicx}	
\usepackage{amsmath}	
\usepackage{adjustbox}
\usepackage[table]{xcolor}

\title[Internal dynamics of Multiple Populations in Globular clusters]{Internal Dynamics of Multiple Populations in 28 Galactic Globular Clusters: A Wide-Field study with Gaia and the Hubble Space Telescope}

\author[G. Cordoni et al.]{
G. Cordoni$^{1,2 }$,\thanks{E-mail: giacomo.cordoni@anu.edu.au}
L. Casagrande$^{1, 2}$,
A.~P. Milone$^{3, 4}$, 
E. Dondoglio$^{3, 4}$,
A. Mastrobuono-Battisti$^{3, 5}$,
S. Jang$^{6}$,
\newauthor
A.~F. Marino$^{4, 7}$,
E.~P. Lagioia$^{8}$,
M.~V. Legnardi$^{3}$,
T. Ziliotto$^{3}$,
F. Muratore$^{3}$,
V. Mehta$^{1}$,
\newauthor
E. Lacchin$^{3, 9, 10, 11}$,
M. Tailo$^{3}$
\\
$^{1}$ Research School of Astronomy and Astrophysics, The Australian National University, Canberra, ACT 2611, Australia \\
$^{2}$ARC Centre of Excellence for All Sky Astrophysics in 3 Dimensions (ASTRO 3D), Australia\\
$^{3}$ Dipartimento di Fisica e Astronomia ``Galileo Galilei'' - Univ. di Padova, Vicolo dell'Osservatorio 3, Padova, IT-35122 \\
$^{4}$ Istituto Nazionale di Astrofisica - Osservatorio Astronomico di Padova, Vicolo dell'Osservatorio 5, Padova, IT-35122 \\
$^{5}$ GEPI, Observatoire de Paris, PSL Research University, CNRS, Place Jules Janssen, 92195 Meudon, France \\
$^{6}$ Center for Galaxy Evolution Research and Department of Astronomy, Yonsei University, Seoul 03722, Korea \\
$^{7}$ Istituto Nazionale di Astrofisica - Osservatorio Astrofisico di Arcetri, Largo Enrico Fermi, 5, Firenze, IT-50125 \\
$^{8}$ South-Western Institute for Astronomy Research Yunnan University, Kunming, 650500, P.R. China \\
$^{9}$ Osservatorio di Astrofisica e Scienza dello Spazio di Bologna, Via Gobetti 93/3, 40129 Bologna, Italy \\
$^{10}$ Institut f\"ur Theoretische Astrophysik, ZAH, Universit\"at Heidelberg, Albert-Ueberle-Straße 2, D-69120, Heidelberg, Germany \\
$^{11}$ INFN - Padova, Via Marzolo 8, I–35131 Padova, Italy
}

\date{Accepted XXX. Received YYY; in original form ZZZ}

\pubyear{2024}

\begin{document}
\label{firstpage}
\pagerange{\pageref{firstpage}--\pageref{lastpage}}
\maketitle

\begin{abstract}
    We present a detailed analysis of the internal dynamics of multiple stellar populations (MPs) in 28 Galactic Globular Clusters (GCs) across a wide field of view, extending from the innermost regions to the clusters' outskirts. Using astro-photometric catalogs from ground-based observations, Gaia, and the Hubble Space Telescope (HST), we identify first- (1P) and second-population (2P) stars, and study the internal dynamics of MPs using high-precision Gaia DR3 and HST proper motions. Our results reveal that while the 1P transitions from isotropy to slight tangential anisotropy toward the outer regions, 2P stars become increasingly radially anisotropic beyond the half-light radius. We also explore the connection between the dynamics of MPs and the clusters' structural and dynamical properties, finding statistically significant differences in the anisotropy profiles of dynamically young and non-relaxed clusters, particularly beyond the 1-2 half-light radii. In these regions, 1P stars transition from isotropic to slightly tangentially anisotropic motion, while 2P stars become more radially anisotropic.
    In contrast, dynamically older clusters, with mixed MPs, exhibit weaker relative differences. Furthermore, clusters with orbits closer to the Galactic center exhibit larger dynamical differences between 1P and 2P stars than those with larger peri-Galactic radii. These findings are consistent with a scenario where 2P stars form in a more centrally concentrated environment, where the interaction with the Milky Way tidal field plays a crucial role in the dynamical evolution of MPs, especially of 1P.
\end{abstract}


\section{Introduction}
\label{sec:intro}

More than two decades of Hubble Space Telescope (HST) observations have revealed that nearly all Galactic Globular clusters (GCs) host Multiple Populations (MPs) with at least two main groups of stars: first-population (1P) and second-population \citep[2P][]{gratton2012, piotto2015, renzini2015, milone2017, bastianlardo2018}. These groups exhibit specific chemical compositions, with 2P stars being for example enriched in Sodium, Nitrogen and Helium  and depleted in Carbon and Oxygen \citep[see e.g.][for recent reviews]{bastianlardo2018,milonemarino2022}. Despite several models have been proposed to explain the formation of MPs, none can simultaneous match the numerous observational constraints \citep[see e.g.][]{renzini2015, bastianlardo2018, milonemarino2022}. Some scenarios predict that 2P stars formed from the ejecta of short-lived massive 1P stars \citep{ventura2001, decressin2007, dercole2010, denissenkov2014}. As the expelled gas accumulated toward the cluster center, 2P stars would dominate the composition of the central cluster regions. Alternatively, the chemical composition of 2G stars could be the result of the accretion of chemically enriched material emitted from 1P stars in binary configuration \citep{demink2009, bastian2013} or supermassive stars \citep{gieles2018}.

While some clusters indeed exhibit a more centrally concentrated 2P, there are numerous examples where 1P and 2P stars are spatially mixed, and even clusters with a more centrally concentrated 1P are observed \citep[see e.g.][]{dalessandro2019, leitinger2023}. Recent studies suggest that the extent of this mixing between 1P and 2P stars may correlate with the dynamical age of the cluster, indicating that dynamically older clusters tend to show a more complete mixing of MPs. Nonetheless, despite the numerous observational constraints, a comprehensive picture of how MPs interact form and evolve over time remains elusive.

In recent years, a promising new path has emerged: the study of the internal kinematics of cluster stars. This approach is crucial for uncovering the physical processes behind the formation of multiple populations. Specifically, \textit{N}-body simulations indicate that the dynamical evolution of 2P stars formed in different environment should differ significantly from that of more spatially extended 1P stars. Such differences may still be detectable in present-day GC kinematics \citep{mastrobuono-battisti2013, mastrobuono-battisti2016, vesperini2013, vesperini2021, henault-brunet2015, tiongco2019, lacchin2022}, provided the populations are not completely relaxed. 

Over the past decade, a number of studies have analyzed the spatial distribution and the internal kinematics of MPs in various GCs, using HST photometry and proper motions \citep[see e.g.][]{libralato2023, leitinger2023}, MUSE/APOGEE line-of-sight (LoS) velocities \citep{szigeti2021, martens2023}, ground-based photometry coupled with Gaia proper motions, or a combination of these methods \citep{milone2018, cordoni2020a, cordoni2020b, cordoni2023, cadelano2024}.
Such works presented the first evidence of significant dynamical differences between 1P and 2P stars in some GCs, qualitatively aligning with $N$-body simulations' results. However, the limited number of clusters and the small field of view prevented a complete and detailed characterization of the phenomenon. More recently, \citet{dalessandro2024} and \citet{leitinger2023} analyzed the 3D kinematics of 16 and 30 GCs, respectively. Both studies combined ground-based photometric data with Gaia proper motions and LoS velocities to investigate the rotation and dynamics of MPs. \citet{dalessandro2024} found differences in the rotation of 1P and 2P stars, with the latter exhibiting stronger rotation. In contrast, \citet{leitinger2023} detected significant rotational differences only in NGC\,0104 (47\,Tuc). \citet{dalessandro2024} also applied analytical models to examine internal dynamics, finding differences in the anisotropy between first-generation and second-generation stars.

In this work we aim to overcome the limitation of previous studies by investigating the internal dynamics of MPs in a sample of 28 GCs, combining photometric information from HST \citep[][]{nardiello2018}, ground-based \citep[see e.g.][]{stetson2019, jang2022} and Gaia \citep{mehta2024} and accurate HST \citep{libralato2022} and Gaia \citep{gaiadr3} proper motions. By doing this we improve the identification of 1P and 2P stars by means of the ChMs, and extend the analyzed field of view from the innermost cluster regions to their outskirt. The paper is structured as follows: in Sec.~\ref{sec:data} we introduce the dataset and MPs selection process. Section~\ref{sec:internal dynamics} presents the study of the internal dynamics in the analyzed clusters and the global dynamical profiles of MPs. Finally, the discussion and summary are presented in Sec.~\ref{sec:discussion} and ~\ref{sec:summary}, respectively.

\section{Data}
\label{sec:data}
To investigate the internal dynamics of MPs in Galactic GCs over a wide field of view, we combined HST photometry \citep{milone2017}, ground-based UBVI photometry \citep{stetson2019,jang2022}, Gaia XP spectro-photometry \citep{mehta2024}\footnote{Gaia XP spectra could be effectively used in only 4 GCs, namely NGC\,0104, NGC\,3201, NGC\,6121 and NGC\,6752.}, with HST \citep{libralato2022} and Gaia DR3 proper motions \citep{gaiadr3}. We refer to the different works for a detailed description of the astro-photometric datasets. The total sample used in this work consists of 28 GCs for which we could reliably select MPs stars over a wide field of view.

To identify 1P and 2P among RGB stars, we have made use of the Chromosome Maps (ChMs) presented in \citet[][]{milone2017} for HST data, photometric diagrams made with the $\cubi$ index\footnote{defined as $\cubi = (U-B) -(B-I)$\citep{milone2012b, monelli2013}}, \citep[][]{cordoni2020a, jang2022} for ground-based observations, and the ChMs introduced in \citet[][]{mehta2024} from Gaia spectro-photometry. ChMs effectively and accurately separate MPs in GCs, with 1P stars defining a clump of stars centered around (0, 0), and 2P stars spread over larger absolute values of $\Delta_\mathrm{BI}$ and  $\Delta_\mathrm{C_{UBI}}$ ($\Delta_\mathrm{F275W, F814W}$ and  $\Delta_\mathrm{C_{F275W, F336W, F438W}}$ in HST filters).  In the next section, we discuss in detail the procedure used to identify 1P and 2P stars in the ground-based ChMs of 28 GCs, whereas we refer to \citet{milone2017} and \citet{mehta2024} for the identification of MPs in HST and Gaia XP photometry. 

The final selection of 1P and 2P stars for each cluster and the analyzed field of view are shown in App.~\ref{app:mpops} as online supplementary material. 

\subsection{Multiple Populations from ground-based photometry}
\label{subsec:mp ground based}
In a nutshell, the $I$ vs. $B-I$ and $I$ vs. $C_\mathrm{UBI}=U - 2B + I$ CMDs of Red Giant Branch stars (RGB) have been used to derive the verticalized $\Delta_\mathrm{BI}$ and  $\Delta_\mathrm{C_{UBI}}$ colors, that are respectively shown on the $x$ and $y$ axis of the ChMs. We refer to \citet{jang2022} for a detailed discussion on how these quantities have been derived. 

\begin{figure*}
    \centering
    \includegraphics[width=0.99\textwidth]{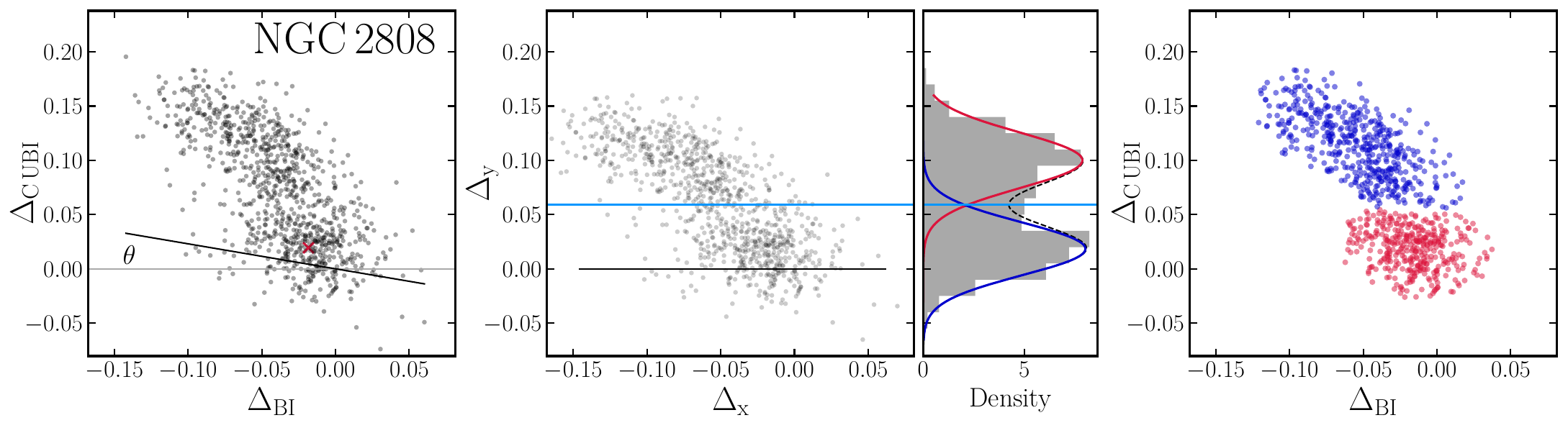}
    \caption{1P and 2P star selection process for the GC NGC\,2808 using ground-based ChM. \textit{Left panel.} Ground-based ChM from \citet{jang2022}. \textit{Middle panel.} ChM rotated by an angle $\theta$ (as shown in the left panel) with a GMM fit applied to $\Delta_\mathrm{y}$. \textit{Right panel.} The same ChM as in the left panel, with 1P and 2P stars highlighted in red and blue, respectively.}
    \label{fig:all chm}
\end{figure*}

To select 1P and 2P stars from the ChMs we adapted the procedure introduced in \citet{milone2017} on HST ChMs to our ground-based ChMs, as done in Jang et al., in review. Figure~\ref{fig:all chm} illustrates the procedure for the cluster NGC\,2808. Briefly, we first selected by eye the bulk of 1P stars among the stars with $(\Delta_\mathrm{BI}, \Delta_\mathrm{C_{UBI}}) \sim (0, 0)$, and fitted a straight line to 1P stars (black solid line in Fig.~\ref{fig:all chm}, leftmost panel). We then rotated the ChMs by the angle $\theta$ defined by slope of the 1P best-fit line. We refer to the new rotated reference frame as $\dx$ and $\dy$ (Fig.~\ref{fig:all chm}, middle panel). By construction, 1P stars define a horizontal distribution in $\dy$ vs $\dx$, and we can use the $\dy$ distribution to disentangle 1P and 2P stars. We used the \texttt{scikit-learn} python package \citep{scikit-learn} and employed Gaussian Mixture Models (GMM) to fit 2 Gaussians to the $\dy$ distribution. We remind here that GMM is applied directly to the data points, so that the results do not depend on the bin choice.
Finally, 1P stars (red points in the rightmost panel) have been selected as stars with $\dy < \dy^\mathrm{sep}$, with $\dy^\mathrm{sep}$ being the intersection between the two best-fit Gaussians, indicated by the horizontal line in the middle panel of Fig.~\ref{fig:all chm}. 2P stars are marked with blue points. Additionally, stars with unusual values of $\dbi$ and $\dcubi$ have been excluded from the selection. On average, less than 1\% of stars have been excluded, and the results are dynamical results are unaffected.

The ground-based ChMs with selected 1P and 2P stars of the 28 analyzed Galactic GCs are shown in App.~\ref{app:mpops} as online supplementary material.

Additionally, we employed the $V$ vs. $\cubi$ pseudo-CMD to extend the regions analyzed in \citet{jang2022}, including stars at larger distance from the cluster centers. We refer to \citet{cordoni2020a} for a detailed description of the selection of MPs.
 
\subsection{Multiple populations from HST photometry}
To study MPs among RGB stars in the innermost cluster regions we exploited the ChMs and selection carried out in \citet{milone2017}. In a nutshell, HST ChMs have been derived by means of the verticalized $m_\mathrm{F814W}$ vs. $m_\mathrm{F275W} - m_\mathrm{F814W}$ (for the abscissa) and $m_\mathrm{F814W}$ vs. $C_\mathrm{F275W, F336W, F438W}$\footnote{$C_\mathrm{F275W, F336W, F438W} = m_\mathrm{F275W} - 2\,m_\mathrm{F336W} + m_\mathrm{F438W}$} CMDs (on the ordinate). 1P and 2P stars have been selected with the procedures detailed in \citet[][see e.g. Sec.~3.3 for a detailed description]{milone2017}. Each HST ChM has then been cross matched with the proper motion catalogs of \citet{libralato2022}.

\subsection{Multiple populations from Gaia XP synthetic photometry}
The recent Gaia DR3, \citep[][]{gaiadr3} publicly released low-resolution spectra, namely XP spectra, for $\sim$200 million sources \citep[][]{deangeli2023}, allowing to derive synthethic photometry in virtually any photometric systems \citep[see e.g.][]{montegriffo2023}. In a recent work, \citet{mehta2024} exploited Gaia XP spectra to compute synthetic photometry special filters designed to maximize the separation between 1P and 2P stars in GCs' RGB. Their analyzed field of view extends well beyond that covered by the ground-based photometric dataset of \citet{stetson2019}, reaching their tidal radius. We refer to \citet{mehta2024} for a detailed description of the procedure used to derive and separate the populations, and for the validation of such identification. In a nutshell, exploiting synthethic spectra with the typical composition of 1P and 2P stars, Mehta and collaborators defined a series of photometric filters to maximize the separation between 1P and 2P stars in 5 Galactic GCs. To validate their MPs identification, \citet{mehta2024} compared the classification with available spectroscopic information, finding consistent results. In this work, we exploit their identification to extend our ground-based dataset and investigate the dynamics of 1P and 2P stars up to the clusters' tidal radii.

\subsection{Anomalous stars \typeii  clusters} \label{subsec:typeii}
Recent studies using HST photometry and high-resolution spectroscopy have shown that about 20\% of Galactic GCs exhibit variations in [Fe/H], \textit{s}-process elements, and total C+N+O \citep[e.g.,][]{marino2015, marino2019, yong2015, milone2017, mckenzie2022}. These clusters are commonly referred to as \typeii, with their heavy-element-enriched stars labeled as ``anomalous''. Photometrically, these clusters show additional sequences in their ChMs, running parallel to the ``canonical'' 1P/2P stars.

Four of the clusters analyzed in this study, namely NGC\,1261, NGC\,1851, NGC\,6934, and NGC\,7089, are classified as \typeii clusters by \citet{milone2017} and \citet{marino2019}. To identify anomalous stars, we used the methodology outlined in \citet{milone2017} on both HST and ground-based ChMs. These stars are highlighted with orange points in Figures~\ref{fig:mpops1}–\ref{fig:mpops3}, and the number of anomalous stars is indicated in Tab.~\ref{tab:tab1}. For NGC\,6934, however, we could only identify anomalous stars using the HST ChM.

\section{Internal dynamics of Multiple Populations}
\label{sec:internal dynamics}
After carefully selecting MPs in the 28 analyzed clusters, we exploited HST \citep{libralato2022} and Gaia DR3 \citep[][]{gaiadr3} proper motions to investigate the internal dynamics of stars belonging to different populations. Specifically, we followed the procedure described in \citet[][for HST]{libralato2022, libralato2023} and \citet{bianchini2018}\footnote{see also \citet{milone2018, bianchini2019, vasiliev2019b, cordoni2020a, cordoni2020b, vasiliev2021, cordoni2023, cadelano2024}.}. We first transformed the celestial coordinates and their proper motion components $(\alpha, \delta, \mua, \mud)$\footnote{$\mua=\mu_\mathrm{\alpha} \cos\delta$} into a Cartesian reference frame $(x, y, \mux, \muy)$, using the orthographic projection \citep[eq. 2 in][]{gaia2018b}, and
then the proper motions into their sky-projected radial and tangential components, defined as $\mur = x \mux + y \muy$ and $\mut = y \mux - x \muy$, with $\mur$ pointing outward (positive) or inward (negative), and $\mut$ positive when counterclockwise. $\mux$ and $\mux$ here indicate the proper motion of each star relative to the motion of the cluster, determined by \citet{vasiliev2021}. Furthermore, considering that the sample of analyzed stars is different from that analyzed in \citet{vasiliev2021}, we repeated the analysis determining the mean cluster motion from our sample of stars, and accounting for systematic errors as in \citet{vasiliev2021}\footnote{routines available at \url{https://github.com/GalacticDynamics-Oxford/GaiaTools}.}, finding consistent results.

The uncertainties on the radial and tangential components of the proper motions have been determined from the uncertainties on $\mua$ and $\mud$, accounting for the correlation between the proper motions. Finally, the radial proper motions have been corrected for the perspective expansion/contraction due the bulk motion of the cluster along the line of sight, as in \citet{bianchini2018}. 

The mean motion and velocity dispersion of 1P and 2P stars have then been computed by dividing the field of view into different concentric annuli containing approximately the same number of stars. For each annulus we determined the mean motion $(\mu_\mathrm{R/T})$ and dispersion $(\sigma_\mathrm{R/T})$ from Gaia DR3 radial and tangential components by minimizing the negative log-likelihood defined in \citet{bianchini2018} including the covariance term as in \citet{sollima2019}. Uncertainties on the determination of the mean motion and dispersion have been determined using Markov Chain Monte Carlo algorithm \citep[\texttt{emcee}][]{foreman2013}. 

The dynamical profiles have been re-computed using different number of bins and  bins with the same radial width to determine the effect of arbitrary bin choice, with consistent results. 
Concerning the HST proper motions, we used the same likelihood without the covariance term, \citep[see e.g. Equation~1 in][]{libralato2022} and maximizing only for the radial/tangential dispersion. This is because HST proper motions are relative to the cluster's bulk motion and do not provide individual stellar absolute mean motion.

In the following sections, we present and discuss the results for selected individual clusters as well as the global dynamical profiles derived by combining data from all clusters. This approach enables us to explore the general dynamical properties of 1P and 2P stars. Our analysis focuses on the dispersion profiles, while we defer a detailed investigation of the rotation of 1P and 2P stars to a forthcoming paper that is currently in preparation.

\subsection{Individual dynamical profiles} \label{subsec:individual}

In this section, we present the study of the dynamical profiles of 1P and 2P stars in a few individual clusters with sufficiently large samples, namely NGC\,0104, NGC\,2808, NGC\,5904, and NGC\,6205. The dynamical profile of the remaining clusters are shown in App.~\ref{app:individual profiles} as online supplementary material. Results for anomalous stars, if present, are shown in orange.

\begin{figure*}
    \centering
    \includegraphics[width=0.24\textwidth]{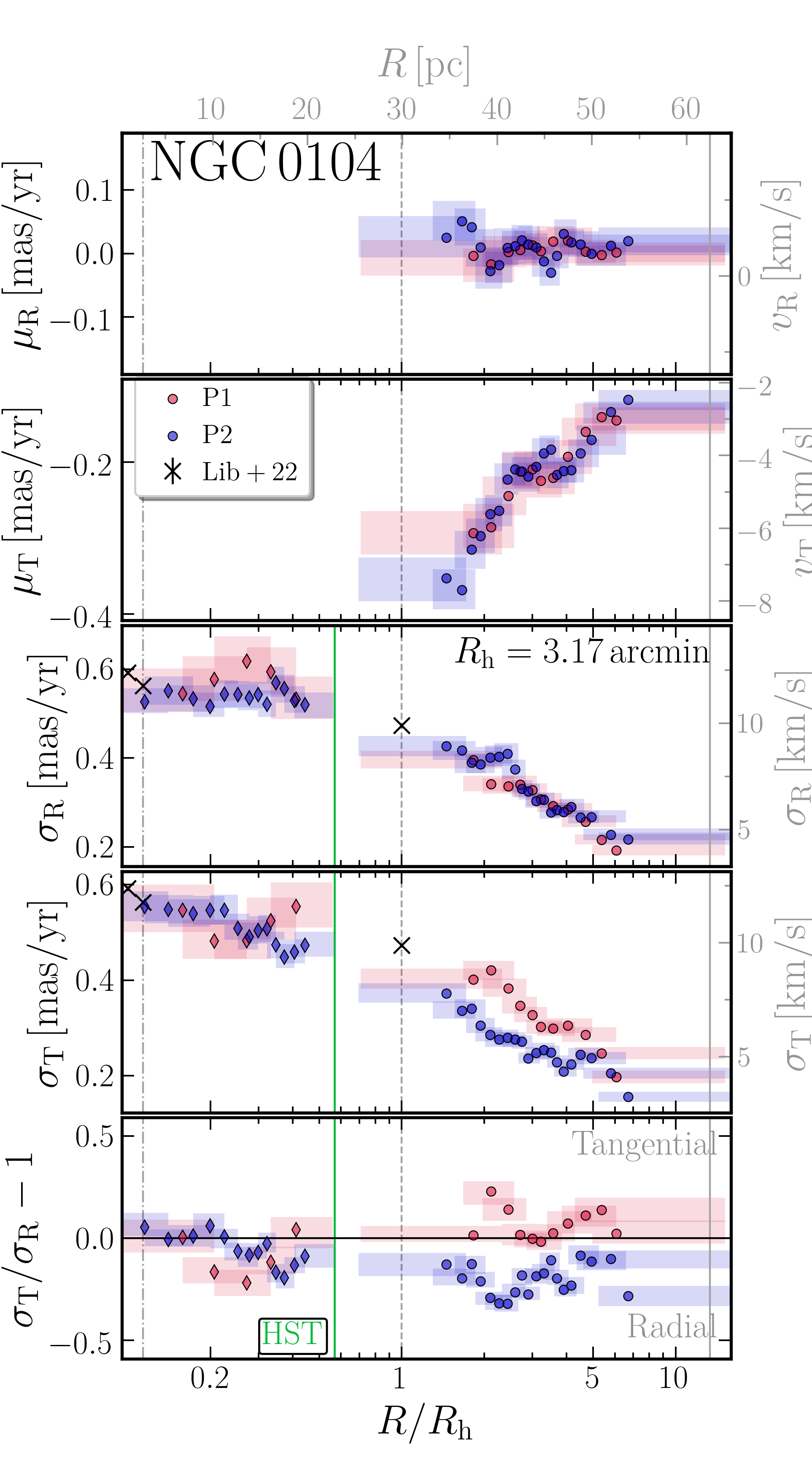}
    \includegraphics[width=0.24\textwidth]{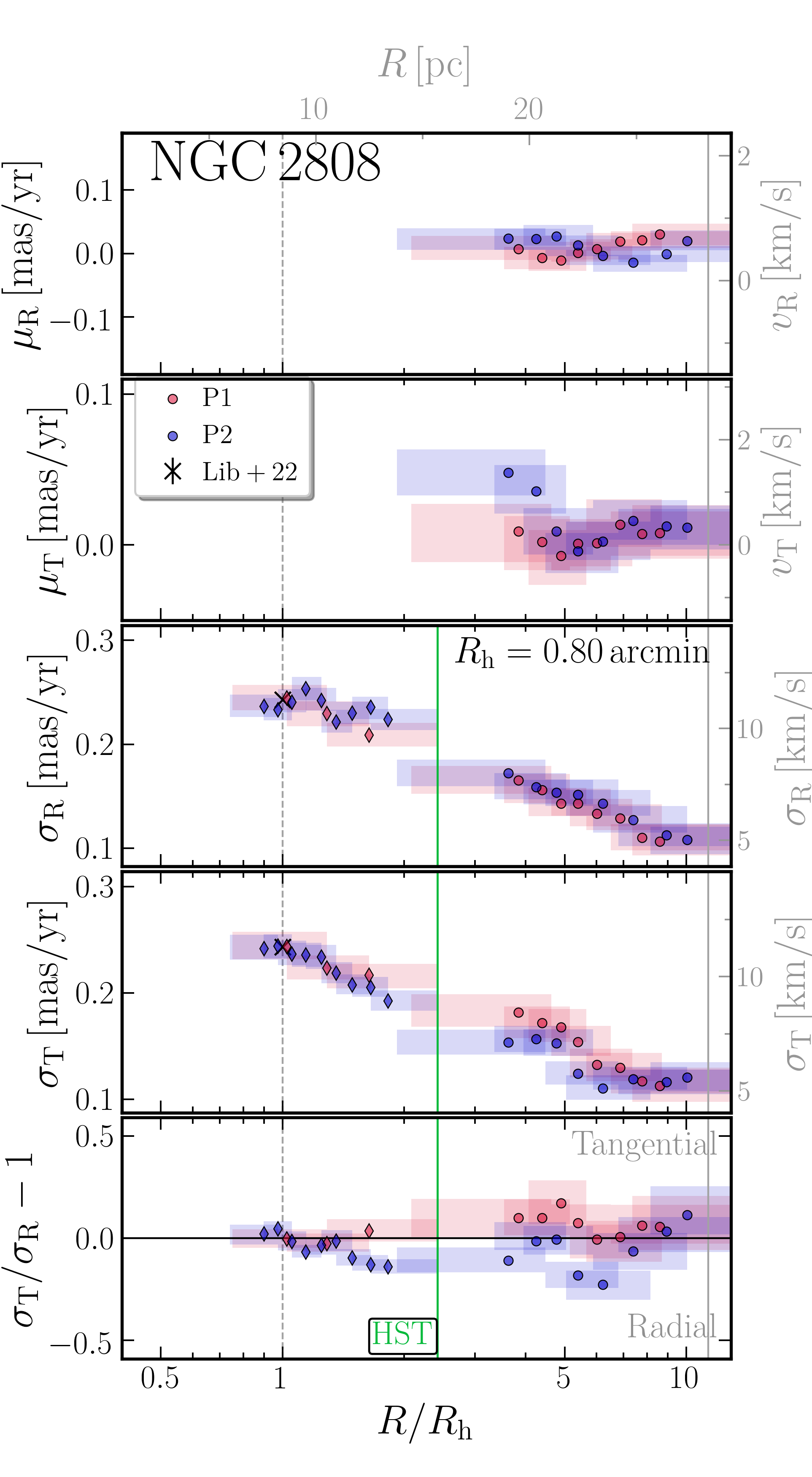}
    \includegraphics[width=0.24\textwidth]{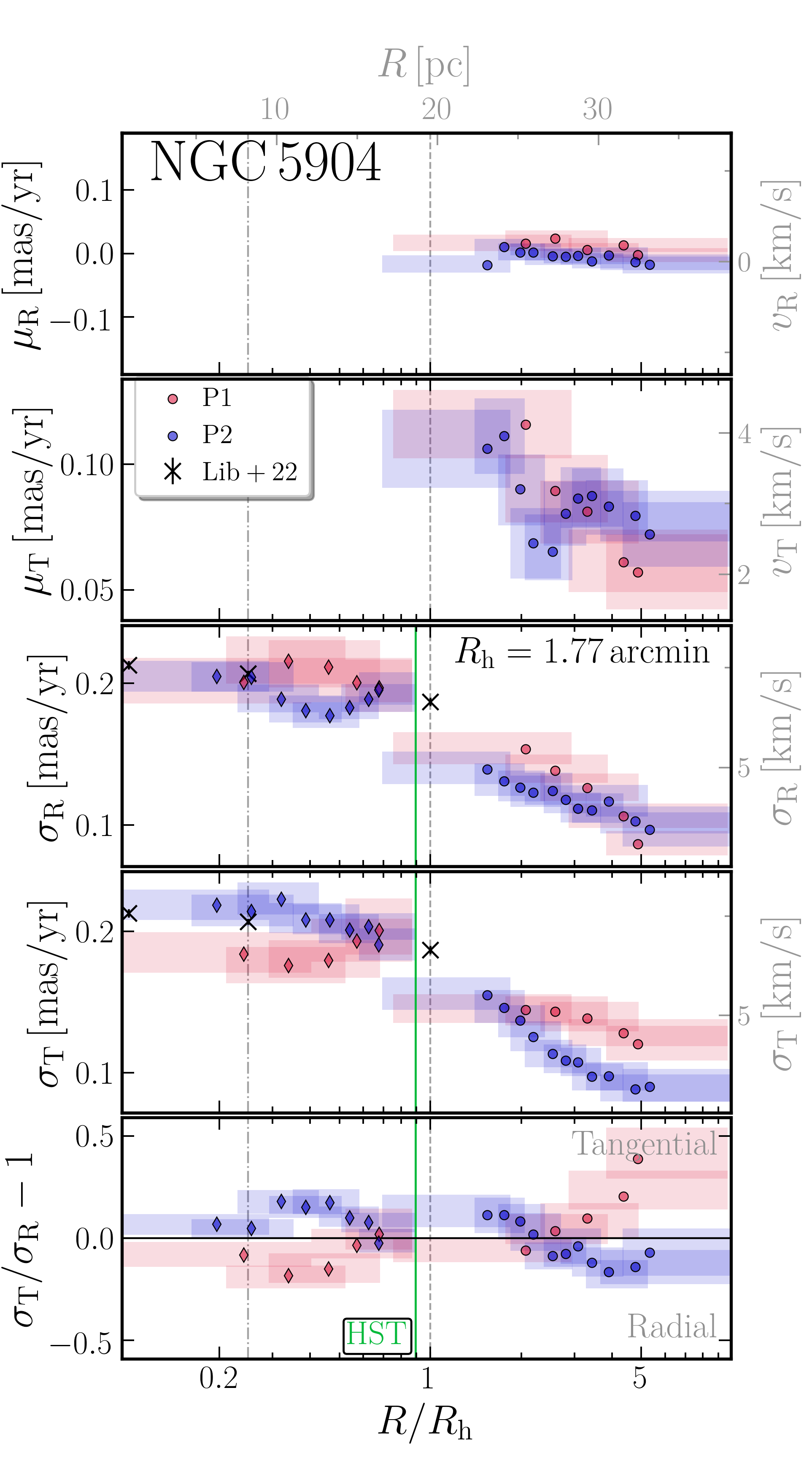}
    \includegraphics[width=0.24\textwidth]{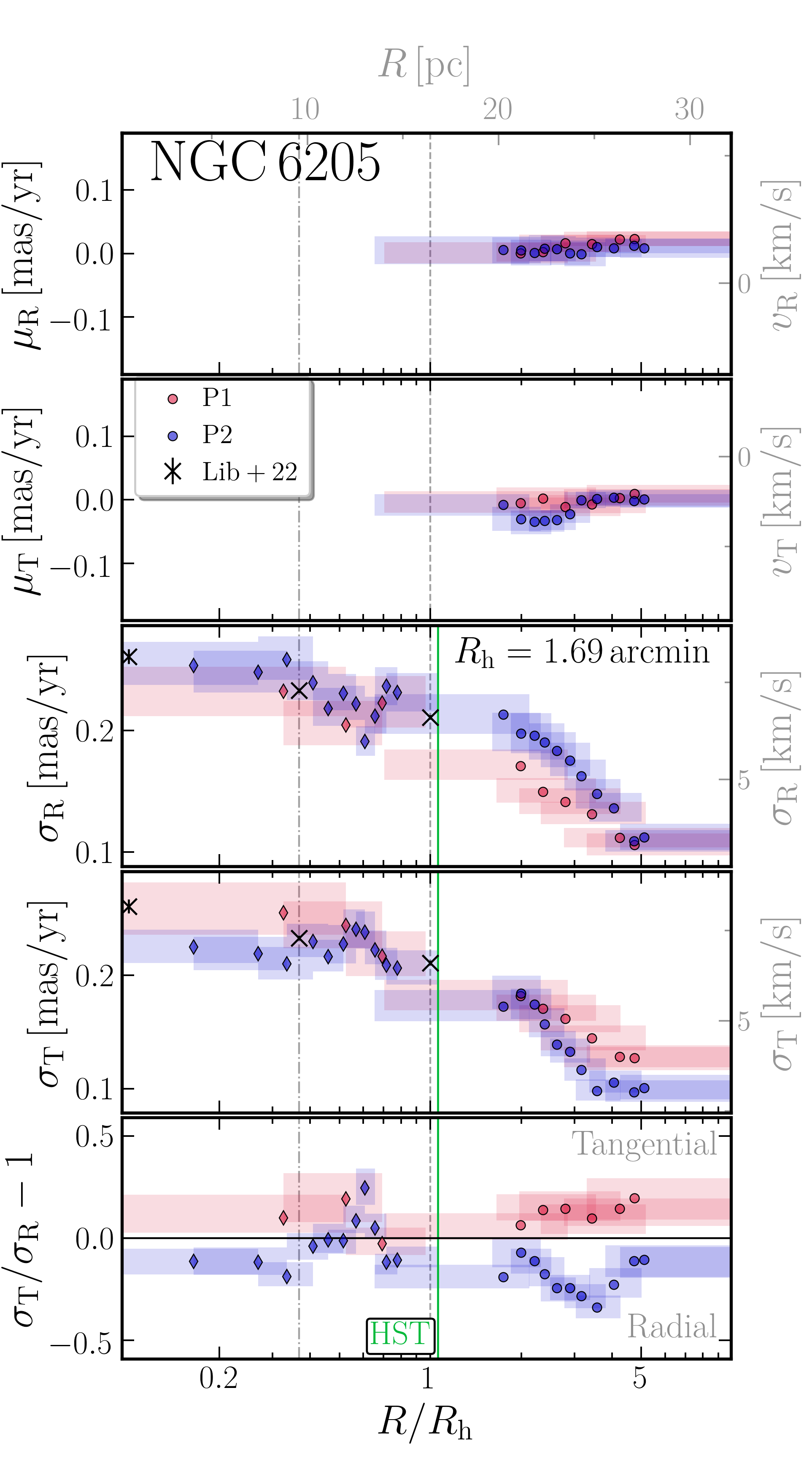}
    \caption{Dynamical profiles of NGC\,0104, NGC\,2808, NGC\,5904 and NGC\,6205. From top to bottom: mean radial and tangential profile, radial and tangential dispersion profiles and anisotropy profile. Red/blue colours refer to 1P and 2P stars respectively, while the corresponding uncertainties are represented by the shaded boxes. The region where HST data, if available, have been used is indicated by the vertical green line, while the dot-dashed and dashed and lines mark the core and half-light radii respectively. Parsec and km/s units are shown respectively in the top and right axis with gray colors.}
    \label{fig:ind prof}
\end{figure*}

Figure~\ref{fig:ind prof} shows, from top to bottom, the mean radial and tangential profiles, the radial and tangential dispersion, and the anisotropy profiles, defined as $\beta=\ani$. Red and blue colours indicate the results for 1P and 2P, determined as described in Sec.~\ref{sec:internal dynamics}, while the shaded rectangles indicate the relative uncertainties and the extension of each radial bin. The distance from the cluster center are in units of $\rh$ indicated in the middle panel \citep[][revision of 2010]{harris1996}. Black diamonds indicate the values of the 1-dimensional dispersion from \citet{libralato2022}, while the regions with available HST data is indicated by the vertical green lines. The top and right axis (shown in gray) of each panel display distances in units of parsec and km/s, converted considering cluster distances determined in \citet{baumgardt2018}.

A visual inspection of the individual profiles reveals that, as expected, the mean radial motions of cluster stars are consistent with 0, as we already subtracted the perspective contraction/expansion. Consistently with \citet{vasiliev2021}, we find that all four clusters exhibit rotation in the plane of the sky. The mean tangential profiles of NGC\,0104 and NGC\,5904 exhibit clear negative and positive values, respectively indicating counterclockwise and clockwise rotation. Possible differences in the rotation profile are consistent with the inferred uncertainties. NGC\,2808 and NGC\,6205, show possible hints of stellar rotation, even though the signal too weak to carry out a comparison. 

Concerning the dispersion profiles, 1P stars exhibit a larger tangential and similar radial dispersion compared with 2P stars, especially outside $\sim 1-2\,\rh$. As a result, 1P stars exhibit an isotropic motion in the central regions, possibly shifting toward tangentially anisotropic motion in the outer regions. On the other hand, the motion of 2P stars is consistent with being isotropic in the cluster center, becoming radially anisotropic beyond 1$\rh$. A similar pattern is also observed in NGC\,3201, consistently with \citet{cadelano2024}. The mean, dispersion and anisotropy profiles of the remaining clusters are displayed in App.~\ref{app:individual profiles}, with anomalous stars in \typeii GCs, namely NGC\,1261, NGC\,1851, NGC\,6934 and NGC\,7089, shown with orange markers\footnote{It is worth noting that the dynamical quantities of anomalous stars are determined from a limited sample of stars in all clusters except for NGC\,1851.}. Overall, we find a good agreement between HST and Gaia dynamical profiles in overlapping or neighboring regions.

The observed dynamical profiles of 1P and 2P stars are qualitatively consistent with the predictions from $N$-body simulations where 2P stars form in a more centrally concentrated environment. We refer to \citet{tiongco2019,vesperini2021, pavlik2022, pavlik2024} for a detailed description of the assumptions adopted in the simulations and the results. Additional discussion is carried out in Sec.~\ref{sec:discussion}.

\subsection{Global dynamical profiles}\label{subsec:stacked}
Given the low number of 1P and 2P stars in many clusters, a detailed analysis of the individual dynamical profiles as a function of cluster radius is often unfeasible. Therefore, similarly to \citet{libralato2023}, we opted to investigate the internal dynamics combining together all clusters. To this goal, we used two different methods: 1) creating stacked 1P/2P catalogs by combining all stars from different clusters, and 2) combining the dispersion profiles of individual clusters to compute global profiles. For the former, the sample was divided into equally populated radial bins, each containing at least 200 stars, to derive global profiles in the same way as individual clusters.

In order to properly compare different clusters, we normalized the radial coordinates to the cluster $\rh$ from \citet[][revision of 2010]{harris1996} and the dispersion profiles to the central dispersion determined in \citet{libralato2022} from HST data. Four clusters do not have determination of the central velocity dispersion from HST data, namely NGC\,1904, NGC\,4147, NGC\,6712 and NGC\,7492. For these clusters, we adopted the central dispersion determined in \citet{vasiliev2021}. We remind here that such approach can be used only to compare dispersion profiles, but not mean motions, as the latter are not on a relative scale. Additionally, combining individual stars from all clusters, we subtracted each cluster’s overall tangential motion profile \citep[derived in][]{vasiliev2021} to account for differences in the rotational pattern of each cluster\footnote{For NGC\,7492, whose internal dynamical profiles were not studied in \citet{vasiliev2021}, we subtracted the cluster's mean tangential motion.}.

In the following, we show the results obtained with approach 1), i.e. combining  individual cluster stars, for 1P and 2P. The profiles derived from stacked 1P/2P profiles, along with supplementary materials and figures, are publicly available on the \href{https://github.com/GiacomoCordoni/streamlitApp_InternalDynamics.git}{GitHub} and as a \href{https://globular-clusters-multiple-populations-dynamics.streamlit.app/}{Streamlit application}. We also include tests, described in the following, on the influence of individual clusters on the global profiles.

\begin{figure*}
    \centering
    \includegraphics[width=0.99\textwidth, trim={0cm 0cm 0cm 1.1cm}, clip]{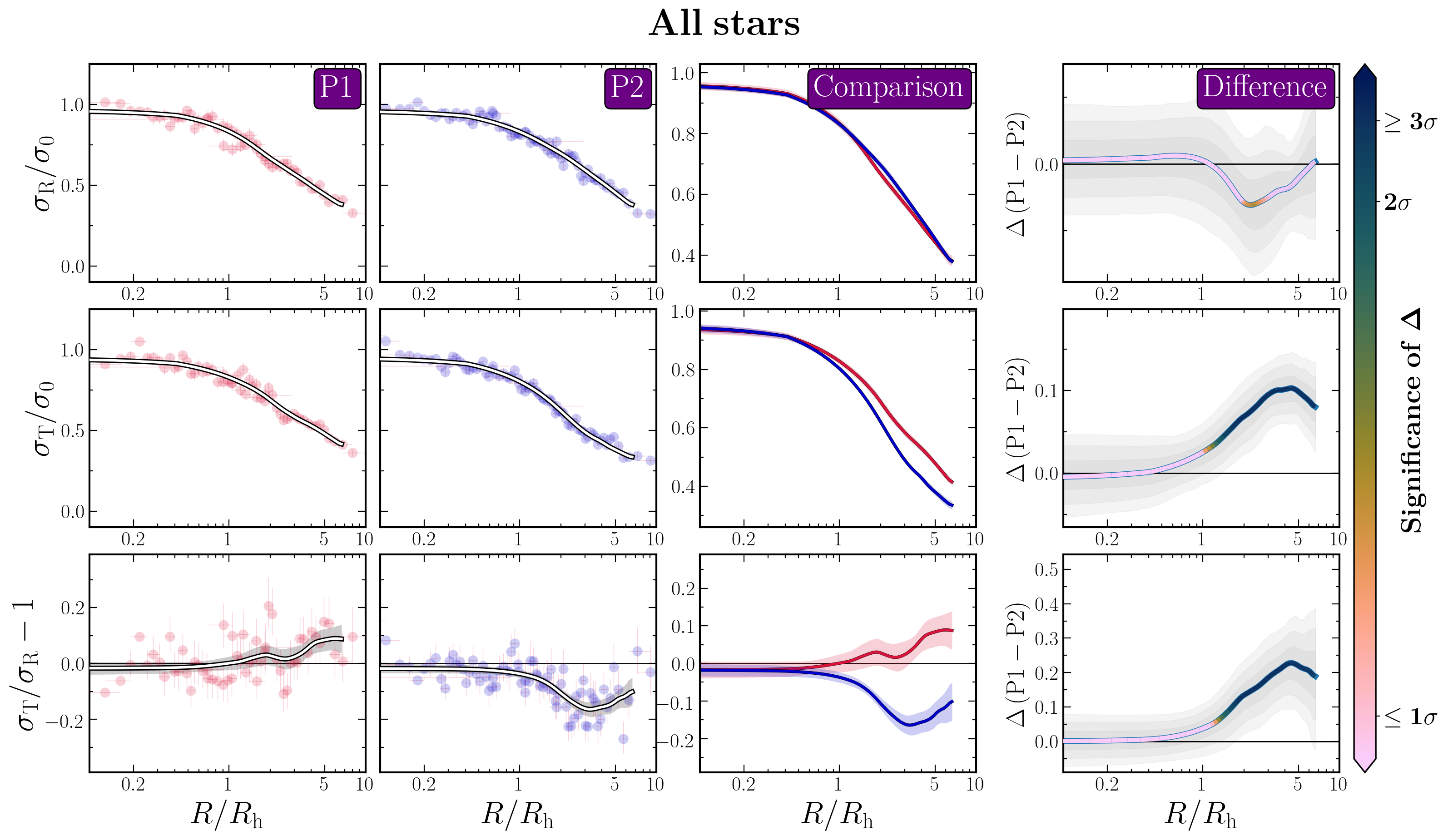}
    \caption{Global dynamical dispersion and anisotropy profiles for the 28 analyzed clusters. Velocity dispersion profiles are normalized to the central dispersion value from \citet{libralato2022}, and radial coordinates are normalized to the clusters’ half-light radius. LOESS trends and uncertainties are represented by solid lines and shaded regions. The first and second columns show the LOESS fit for 1P and 2P separately, while the third column displays their comparison. The fourth column presents the difference between the 1P and 2P global profiles along with its statistical significance. The 1, 2, and 3 $\sigma$ confidence intervals are indicated by gray shaded regions, and the significance of the difference is represented by the color of the lines. Points with significance lower than $1\sigma$ are shown in the same color for clarity.}
    \label{fig:stacked all}
\end{figure*}

The normalized global dispersion and anisotropy profiles as a function of distance from cluster center for the 28 analyzed GCs are shown in Fig.~\ref{fig:stacked all}. Global profiles for 1P and 2P stars are shown in the first and second columns, while the comparison between the average trend is shown in the third column. The global profiles were determined using the Locally Weighted Regression (LOESS) as implemented in \citet{cappellari2013}. To estimate the uncertainties in the average trend, we repeated the LOESS fitting on 1000 bootstrapped samples, drawing the values of $\sigma_\mathrm{R/T}$ in each realization from a Gaussian distribution centered on the observed value with a dispersion equal to the observed uncertainties. The uncertainties were determined as the 16$^\mathrm{th}$ and 84$^\mathrm{th}$ percentiles of the 1000 LOESS fits. This is a more robust approach than simply bootstrapping the sample, as it accounts for observed uncertainties as well. Finally, we estimated the significance of the observed differences between 1P and 2P stars accounting for uncertainties in both data and fitting. The details of the procedure are outlined in App.~\ref{app:significance}. In a nutshell, for each of the 1000 realizations of the LOESS fit, we computed the difference between 1P and 2P as a function of the radial coordinate. The 1, 2 and 3$\sigma$ confidence regions are shown as gray shaded regions in the rightmost panels, where we display the difference between 1P and 2P a function of normalized radius. Additionally, we quantified the statistical significance of each point as the fraction of simulations which returned a difference smaller than the observed one, at the same radial location. The color of the line in the rightmost panels is indicative of of the statistical significance, in units of $\sigma$, as indicated in the right colorbar.

While there are no differences between the global radial dispersion profiles of 1P and 2P stars, 2P stars exhibit a lower tangential dispersion compared to 1P stars, especially in the outer regions (e.g., $R > 1-2\rh$). Consequently, the 1P displays isotropic or tangentially anisotropic motion in these regions, while the 2P is radially anisotropic. The observed differences are significant beyond $3\sigma$ between 2 and $\sim 6\,\rh$. Our global profiles within $\rh$ are also consistent with the conclusions of \citet{libralato2023}. Additionally, although we computed the global profile of anomalous stars, $\sim 70$\% of these stars belong to NGC\,1851, making the profile unrepresentative of the entire cluster sample. Therefore, we choose not to present it. Nevertheless, we note that the profile is qualitatively consistent with 2P global profiles. 

To identify and investigate potential biases in the global profiles caused by certain clusters contributing more to the stacked datasets (e.g., the most populated clusters), we conducted the following tests: \textit{i)} derived global profiles using only the six most populated clusters (NGC\,0104, NGC\,2808, NGC\,5272, NGC\,6205, NGC\,5904, and NGC\,7089); \textit{ii)} repeated the analysis excluding these six clusters; and \textit{iii)} randomly sampled an equal number of 1P and 2P stars (50 stars, or the total number of available stars for clusters with fewer than 50 1P/2P) from each cluster to create 1,000 combined dataset realizations. In all three tests, we obtained results  consistent with those shown in Fig.~\ref{fig:stacked all} (i.e. radially anisotropic 2P and isotropic/tangentially anisotropic 1P), with the main difference being the increased uncertainties resulting from smaller samples. Furthermore, for each stacked dataset, we generated heat maps to illustrate the contribution of each cluster in each radial bin and found no significant biases. 
To further validate our findings, we recomputed the global profiles from the stacked individual dynamical profiles (e.g. the profiles presented in Figs.~\ref{fig:raddis}-\ref{fig:beta}), again obtaining consistent results. The results of these tests and their descriptions are available on \href{https://github.com/GiacomoCordoni/streamlitApp_InternalDynamics.git}{GitHub} and as a \href{https://globular-clusters-multiple-populations-dynamics.streamlit.app/}{Streamlit application}.

\section{Discussion}\label{sec:discussion}
In the last decade, several theoretical works based on $N$-body simulations have investigated the dynamical evolution of MPs in GCs and their potential for distinguishing between different formation scenarios. The velocity dispersion and anisotropy profiles of MPs can provide fundamental insights into the origins of these stellar populations \citep[see e.g.][]{henault-brunet2015, mastrobuono-battisti2013, mastrobuono-battisti2016, mastrobuono-battisti2021, tiongco2019, vesperini2021, lacchin2022, hypki2022, pavlik2022, pavlik2024}. However, due to dynamical evolution and spatial mixing, possible differences in the internal dynamics of MPs disappear as the clusters reach relaxation. As shown in Fig.\ref{fig:ind prof} for individual clusters and Fig.\ref{fig:stacked all} for all clusters, the tangential dispersion and anisotropy profiles of 1P and 2P stars exhibit some differences, with 1P being isotropic across the entire cluster field and 2P stars shifting from isotropy to radially anisotropy beyond $\rh$.

\begin{figure*}
    \centering
    \includegraphics[width=0.99\textwidth, angle=00]{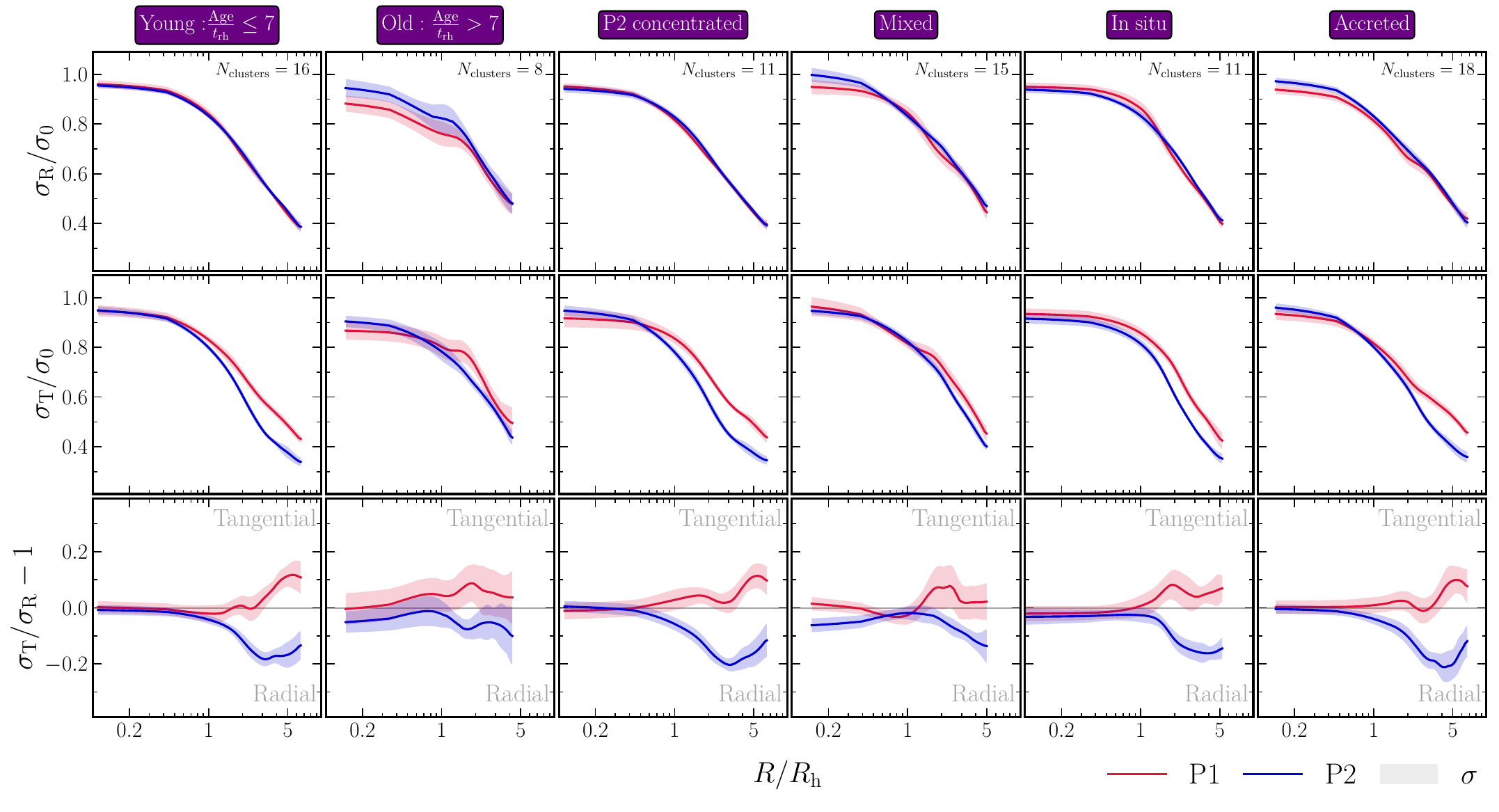}
    \includegraphics[width=0.99\textwidth, angle=00]{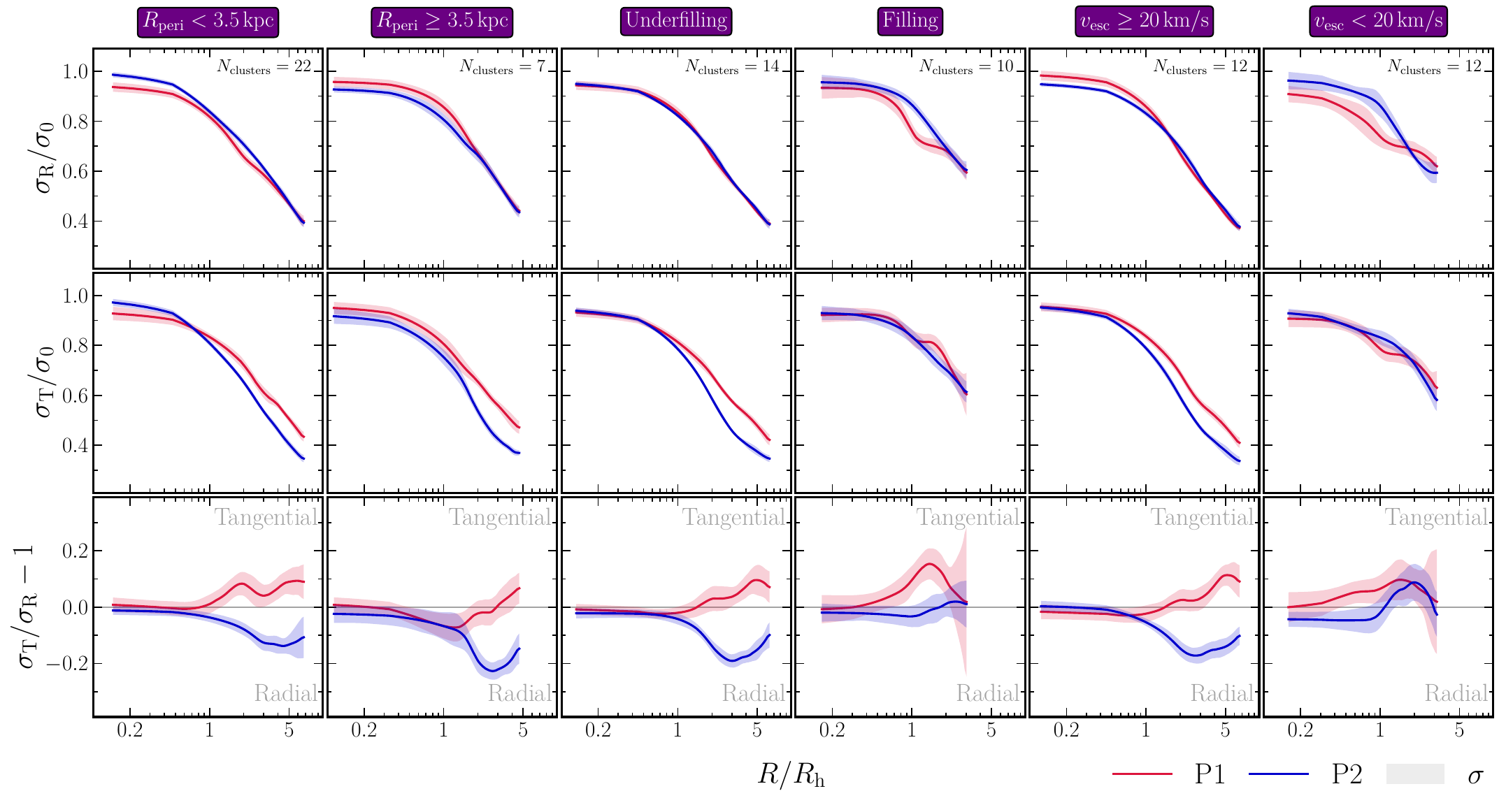}
    \caption{Comparison between dispersion and anisotropy profiles of clusters with different properties, as indicated in the insets. Details about each group and the median trend is discussed in the text. Red and blue colors indicate 1P and 2P stars. The statistical significance of the difference between 1P and 2P trend is shown in App.~\ref{app:significance}.}
    \label{fig:comparison}
\end{figure*}

To investigate how these dynamical differences depend on the clusters' properties evolutionary state, we divided our sample of 28 clusters into different groups based on various internal and external factors: \texttt{dynamical ages} 
\citep[determined as the ratio between cluster age and half-mass relaxation time, see e.g.][]{libralato2022}, \texttt{tidal filling factor} defined as $\mathcal{R}=\rh/\rj$ \citep{baumgardt2010, shin2013} with $\rj$ being the Jacobi radius derived in \citet{balbinot2018}\footnote{Clusters with $\mathcal{R} < 0.05$ are considered in ``isolated regime'' or tidally underfilling, while clusters with $\mathcal{R} > 0.05$ are in ``tidal regime'', or tidally filling.}, 
\texttt{spatial mixing of MPs} \citep[as determined in][and Jang et al., in review]{leitinger2023}, \texttt{tidal interaction} with the Milky Way \citep[determined on the basis of the peri-Galactic radius, see e.g.][]{milone2020, libralato2023}\footnote{Consistently with literature works, we divide clusters with $R_\mathrm{peri}\lessgtr3.5\,\mathrm{kpc}$}, \texttt{clusters' origin} (e.g. in situ or accreted) as identified in \citet{massari2019} and \texttt{escape velocity} \citep[$\vesc$, see e.g.][]{baumgardt2018, mastrobuono-battisti2021}\footnote{See e.g. the analysis of \citet{baumgardt2018} and \citet{mastrobuono-battisti2021} and the connection between the cluster's $\vesc$ and MPs. The value of the limiting escape velocity, i.e. $20\,\mathrm{km/s}$, is determined examining the distribution of escape velocities from \citet{baumgardt2018}.}. The groups are shown in the table provided in App.~\ref{app:significance}. We find worth mentioning that different groups have overlap, as some clusters' properties are not independent one from the other \citep[e.g. the spatial mixing of 1P and 2P is connected with the cluster dynamical age][]{dalessandro2019, leitinger2023}. Additionally, we also divided clusters based on their \texttt{internal rotation} \citep[as determined in][]{vasiliev2021}, but, to keep the figure simpler, we do not show the results in Fig.~\ref{fig:comparison}. Nonetheless, the difference between 1P and 2P anisotropy profiles for the groups, along with the statistical significance, is included in Fig.~\ref{fig:significance1}.

The dynamical profiles for different groups of clusters are displayed in Fig.~\ref{fig:comparison}. Each group is indicated in the top inset, together with the number of clusters in each group. The global profiles (solid lines) have been computed with the LOESS algorithm, while the uncertainties (shaded regions, 1$\sigma$) have been determined by bootstrapping with replacements a 1000 times and accounting for the observational uncertainties (see Sec.~\ref{subsec:stacked} for a detailed description). The significance of the differences between the anisotropy profiles of 1P and 2P for each group is presented in App.~\ref{app:significance}, with the description of the procedure adopted to estimate it.

We find significant differences, i.e. above the 3$\sigma$ level, in the average anisotropy profiles of many of the analyzed groups. Overall, we find that the 2P is more radially anisotropic beyond $\rh$, while 1P is isotropic and become slightly tangentially anisotropic beyond 2-3$\rh$. There are important differences in the 1P-2P relative dynamics among some of the analyzed groups. For example, while dynamically young clusters and those with centrally concentrated 2P stars show clearly different dispersion and anisotropy profiles, these differences are less pronounced in intermediate-age and mixed-population clusters. Indeed, dynamical profile of 1P and 2P in dynamically old clusters are consistent within $1\sigma$. These findings are qualitatively consistent with the simulations by \citet[][see, e.g., their Fig.~9]{tiongco2019} and \citet[][see, e.g., their Figs.~13-17]{vesperini2021}, which suggest that due to different initial spatial configurations (such as centrally concentrated 2P stars) or initial isotropy, 1P stars tend to develop isotropic motion, while 2P stars display more radially anisotropic motion, especially in the outer regions. Over time, these dynamical differences gradually diminish as spatial and dynamical mixing of MPs occurs. 

The comparison between clusters with small and large peri-Galactic radii (referred to as ``inner'' and ``outer'' clusters, respectively) reveals distinct dynamical patterns (Fig.~\ref{fig:comparison}, bottom row). In inner clusters, 1P stars exhibit tangential anisotropy beyond $\rh$, while in outer clusters, 1P is either isotropic or slightly radially anisotropic between 1 and 2 $\rh$. In both cases, 2P stars become increasingly radially anisotropic as they move outward. Although the differences in 1P dynamics between inner and outer clusters are only marginally significant in 1-2$\rh$ region (ranging between 1–2$\sigma$), these findings suggest that the Milky Way’s tidal field significantly influences the dynamical evolution of MPs. Moreover, in clusters with small peri-Galactic radii, 1P stars appear to behave like tidally filling systems, where the outer regions develop slight tangential anisotropy, whereas 2P stars resemble tidally underfilling systems with radial anisotropy. In contrast, weaker interactions with the Milky Way’s tidal field result in a more isotropic 1P population. This scenario aligns qualitatively with the theoretical models presented in \citet{tiongco2019, hypki2022, pavlik2024}.

To further explore this hypothesis, we also examined the internal dynamics of MPs in clusters classified as tidally underfilling ($\mathcal{R} < 0.05$) and tidally filling ($\mathcal{R} > 0.05$). While tidally underfilling clusters exhibit relative differences consistent with other groups, such as isotropic 1P and radially anisotropic 2P, MPs in tidally filling clusters display a different pattern. Specifically, 1P stars are tangentially anisotropic, while 2P stars exhibit isotropic motion across the entire field. Interpreting these results is challenging, as the tidal filling factor $\mathcal{R}$ is closely linked to other properties, such as $\vesc$, $R_\mathrm{peri}$, and cluster mass. Specifically, tidally underfilling clusters tend to be more massive, with larger $\vesc$ and smaller $R_\mathrm{peri}$.

A qualitatively similar pattern is observed in clusters with low and high escape velocities, with the latter often exhibiting tangentially anisotropic 1P stars. This difference may be due to the preferential loss of 1P stars on radial orbits in clusters with higher $\vesc$ \citep[see, e.g.,][]{mastrobuono-battisti2021, lacchin2024}. However, the relationship between $\vesc$ and the fraction of MPs remains complex and is highly dependent on the specific formation scenario adopted for MPs.

1P and 2P stars exhibit comparable relative differences in clusters, regardless of whether they are in situ or accreted, and whether they are rotating or non-rotating (see, e.g. App.~\ref{app:significance}). In these cases, 2P stars remain radially anisotropic, while 1P stars display isotropic motion across the entire cluster field.

In summary, the results presented in Fig.~\ref{fig:comparison} indicate that clusters in a less dynamically evolved state show significant dynamical differences among MPs. Additionally, the interaction with the Galaxy appears to play a crucial role in shaping the evolution of different populations. These findings align with the conclusions of \citet{libralato2023} regarding the internal dynamics of MPs in the innermost regions. However, while our  results clearly support the presence of a common dynamical pattern, it is important to note that a small number of clusters deviate from this trend, exhibiting distinct dynamical properties. The nature of these differences remains unclear and may be influenced by specific cluster properties and their evolutionary histories.

\subsection{Comparison with the literature} \label{subsec:comparison}
In recent years, numerous studies have investigated the dynamics of MPs in GCs (see references in Sec.~\ref{sec:intro}). Here, we provide a comparison with the findings of \citet{libralato2023, dalessandro2024, leitinger2024}, who analyzed 56, 16 and 30 GCs, respectively. Our conclusions regarding the dynamical properties of 1P/2P stars in both dynamically young and old clusters are consistent with the results of \citet{libralato2023} and \citet{dalessandro2024}. Specifically, our findings support the presence of a general trend in the dynamical properties of 1P and 2P stars, which suggests a common underlying dynamical pattern. On a cluster-by-cluster basis, our anisotropy profiles qualitatively align with the $\alpha_\beta$ parameters reported by Dalessandro et al., with the exception of NGC\,5927. However, direct comparisons are limited, as their study employs analytical fits to model the dispersion and anisotropy profiles, whereas we utilize profiles derived from radial bins. Similarly, we cannot directly compare results on the internal rotation of individual clusters or the global rotation profile presented by \citet{dalessandro2024}, as we chose not to fit the tangential profiles with analytical models. A direct comparison with \citet{leitinger2024} is also challenging, as they focus more on MPs rotation and present anisotropy profiles for all cluster stars  without separating MPs.

Finally, the individual cluster profiles derived in this work are qualitatively consistent with the results of \citet{milone2018, cordoni2020a, cadelano2024}.

\section{Summary and conclusion}\label{sec:summary}

In this study, we present a detailed investigation of the internal dynamics of multiple populations across a wide field of view, from the innermost arcmin to the clusters' outskirts, in a large sample of 28 Galactic GCs. Using HST, ground-based and XP synthethic photometry, we identified first- and second-population stars. To achieve this, we exploited the pseudo CMD dubbed Chromosome maps as well as the $\cubi$ index, where 1P and 2P stars form well-separated groups. The internal dynamics of MPs was investigated using HST and Gaia DR3 proper motions, allowing us to determine the mean, dispersion, and anisotropy profiles as function of distance from the cluster center. The analyzed cluster regions range from 0.2$\rh$ to more than 10 $\rh$. The analysis presented in this work confirms and extends the results of previous studies based on HST and Gaia proper motions \citep[see e.g.][]{richer2013, bellini2015, bellini2018, milone2018, cordoni2020a, cordoni2020b, libralato2023, cadelano2024, dalessandro2024}. 

Our analysis of individual clusters reveals distinct dynamical differences in the anisotropy profiles of 1P and 2P stars in several cases. On average, 2P stars are more radially anisotropic beyond $\rh$, whereas 1P stars generally exhibit isotropic motion, with some showing tangential anisotropy beyond 2-3$\rh$. These anisotropy differences between 1P and 2P stars are primarily driven by a lower tangential dispersion of 2P, with no significant differences observed in the radial component. These results agree with the analysis of \citet{libralato2023} for the innermost cluster regions.

Studying the global trends in dispersion and anisotropy profiles, derived by combining individual clusters' catalogs (see Sec.~\ref{subsec:stacked} for a description of the normalization), we observe significant differences between 1P and 2P stars, especially outside $\rh$ where the differences are significant beyond the 3$\sigma$ level.

In dynamically young clusters, 1P stars are isotropic in the inner regions but become slightly tangentially anisotropic toward the outskirts. In contrast, 2P stars are isotropic in the cluster centers and become radially anisotropic in the outer regions.  This result is in agreement with the results presented in \citet{dalessandro2024} for a sample of 16 GCs. These patterns are also evident in clusters where 2P stars are more centrally concentrated and in clusters with escape velocities exceeding $20\,\mathrm{km/s}$. However, dynamical evolved clusters, with spatially mixed MPs and lower escape velocities do not exhibit these dynamical differences. Rotating/non-rotating clusters, and in situ/accreted clusters show similar relative differences between 1P and 2P stars.

We also explore the influence of the Milky Way’s tidal field on the dynamical properties of MPs by analyzing clusters with different peri-Galactic radii. Our findings reveal a possible connection between the dynamical behavior of 1P stars and the strength of the Milky Way’s tidal field. In clusters with orbits closer to the Galactic center, where the tidal field is stronger, 1P stars tend to exhibit tangential anisotropy beyond 1-2$\rh$. Conversely, clusters with weaker interactions with the Milky Way’s tidal field, 1P stars display isotropic motion. This suggests that the Milky Way’s tidal field plays an important role in the dynamical evolution of MPs. Further supporting this conclusion, our analysis shows that tidally underfilling and filling clusters exhibit distinct relative patterns in the dynamical profiles of 1P and 2P stars. 

The observed differences in the internal dynamics between 1P and 2P stars qualitatively align with the predictions of $N$-body and theoretical simulations by \citet{mastrobuono-battisti2016, tiongco2019, vesperini2021, pavlik2024}. These studies indicate that the distinct dynamical properties of 1P and 2P stars are indicative of 2P stars forming in a more centrally concentrated environment. 

Overall, the analysis presented in this paper offers key insights into the formation scenarios of multiple stellar populations and their relationship with both internal factors (such as $\vesc$ and dynamical age) and external influences (such as interaction with the Milky Way’s tidal field).

\section*{Acknowledgments}

SJ acknowledges support from the NRF of Korea (2022R1A2C3002992, 2022R1A6A1A03053472). EPL acknowledges support from the ``Science
\& Technology Champion Project'' (202005AB160002)
and from the ``Top Team Project'' (202305AT350002),
all funded by the ``Yunnan Revitalization Talent Support Program''. TZ acknowledges funding from the European Union’s Horizon 2020 research and innovation program under the Marie Sk\l odowska-Curie
Grant Agreement No. 101034319 and from the European Union – Next Generation EU. This work has received funding from  ``PRIN 2022 2022MMEB9W - \textit{Understanding the formation of globular clusters with their multiple stellar generations}'' (PI Anna F.\,Marino),  and from INAF Research GTO-Grant Normal RSN2-1.05.12.05.10 -  (ref. Anna F. Marino) of the ``Bando INAF per il Finanziamento della Ricerca Fondamentale 2022''. EL acknowledges financial support from the European Research Council for the ERC Consolidator grant DEMOBLACK, under contract no. 770017. AMB acknowledges funding from the European Union’s Horizon 2020 research and innovation programme under the Marie Sk\l{}odowska-Curie grant agreement No 895174.

\section*{Data Availability}

Relevant data underlying this work is available in the article. Additional figures and tests mentioned in the artcile are available on \href{https://github.com/GiacomoCordoni/streamlitApp_InternalDynamics.git}{GitHub} and as a \href{https://globular-clusters-multiple-populations-dynamics.streamlit.app/}{Streamlit application}. All other data will be shared upon reasonable request to the corresponding author. 

\bibliographystyle{mnras}
\bibliography{main} 


\appendix

\section{Selection of Multiple Populations}\label{app:mpops}
\begin{figure*}
    \centering
    \fbox{\includegraphics[width=0.48\textwidth]{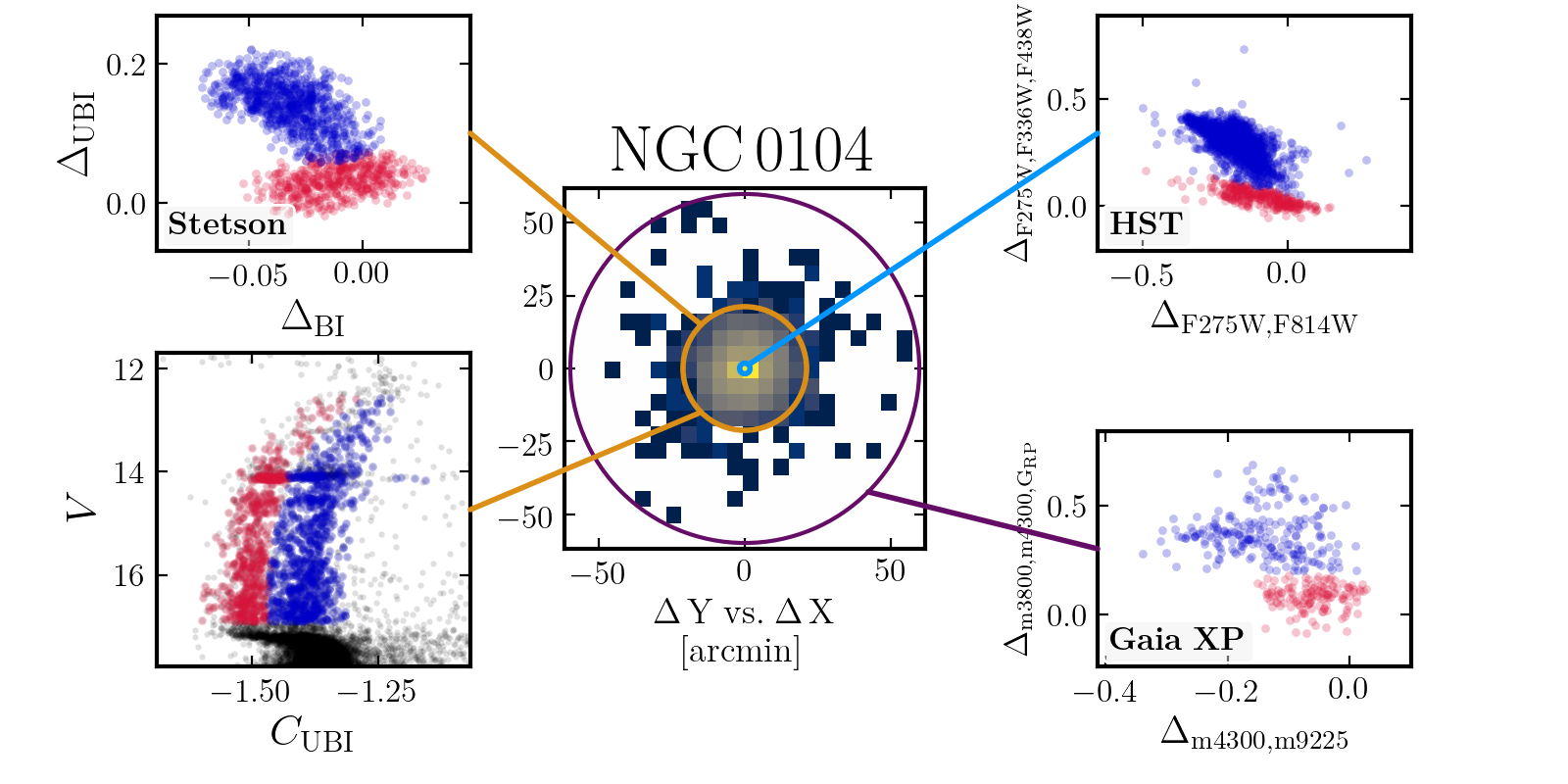}}
    \fbox{\includegraphics[width=0.48\textwidth]{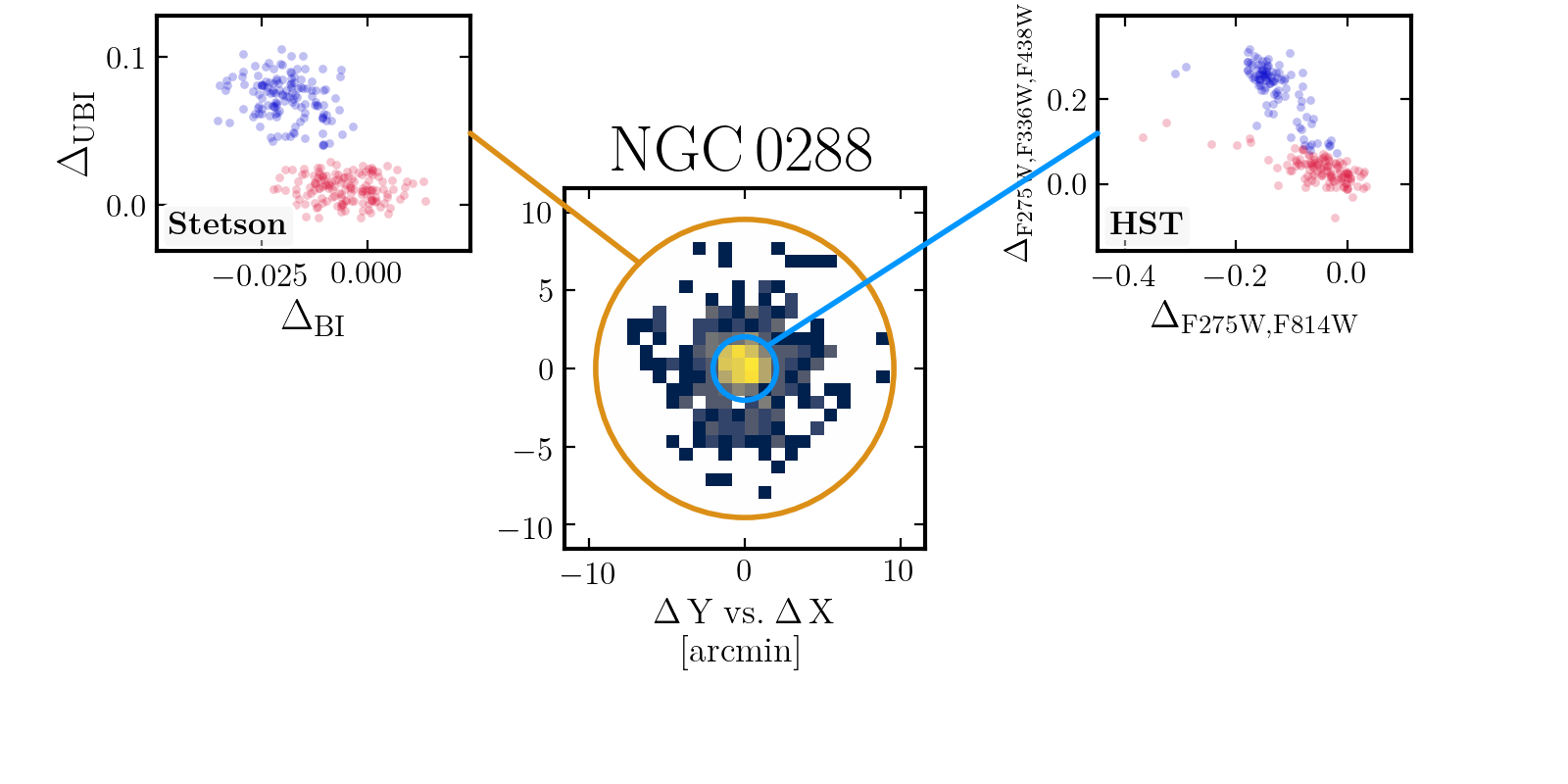}}
    \fbox{\includegraphics[width=0.48\textwidth]{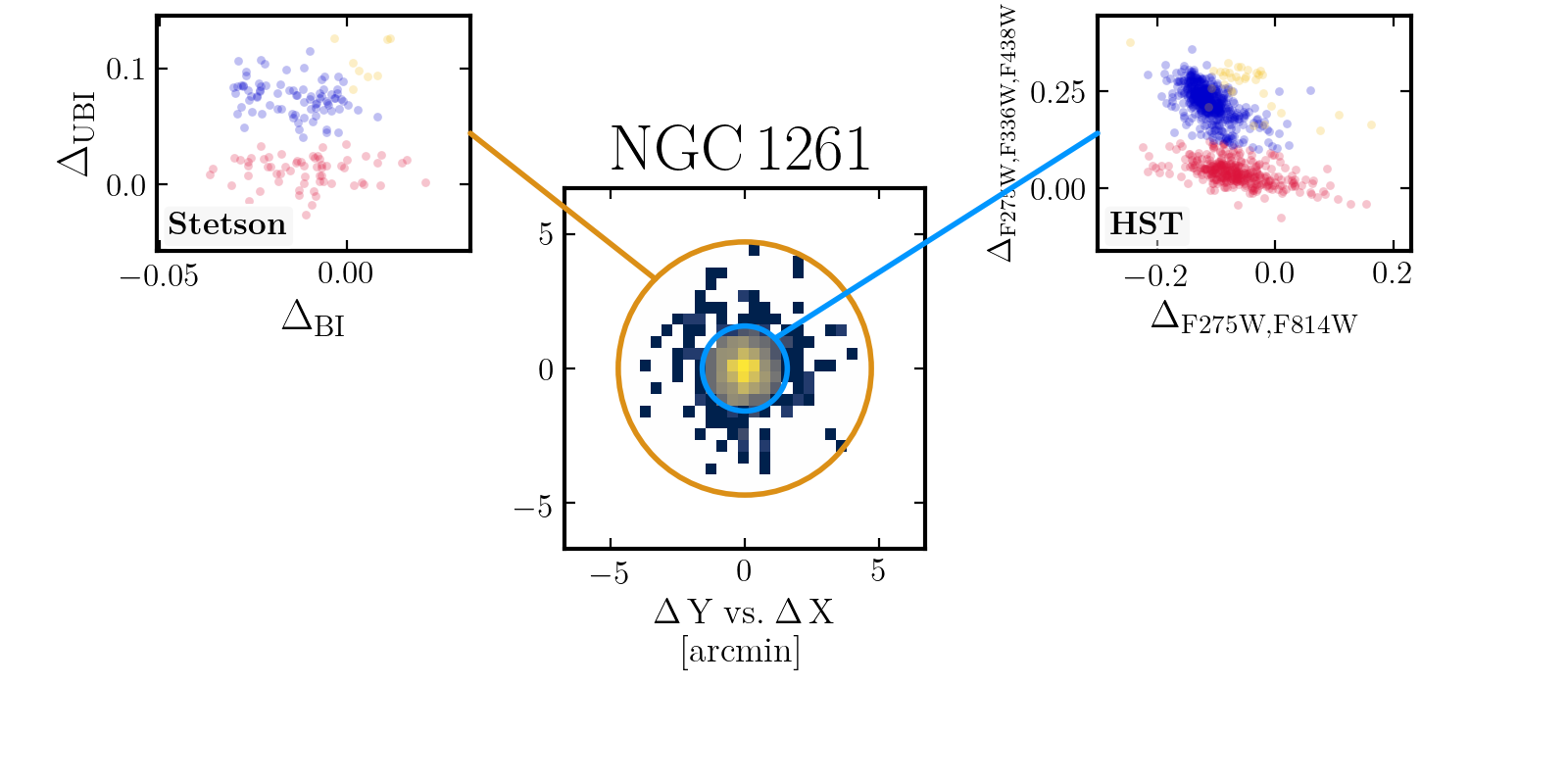}}
    \fbox{\includegraphics[width=0.48\textwidth]{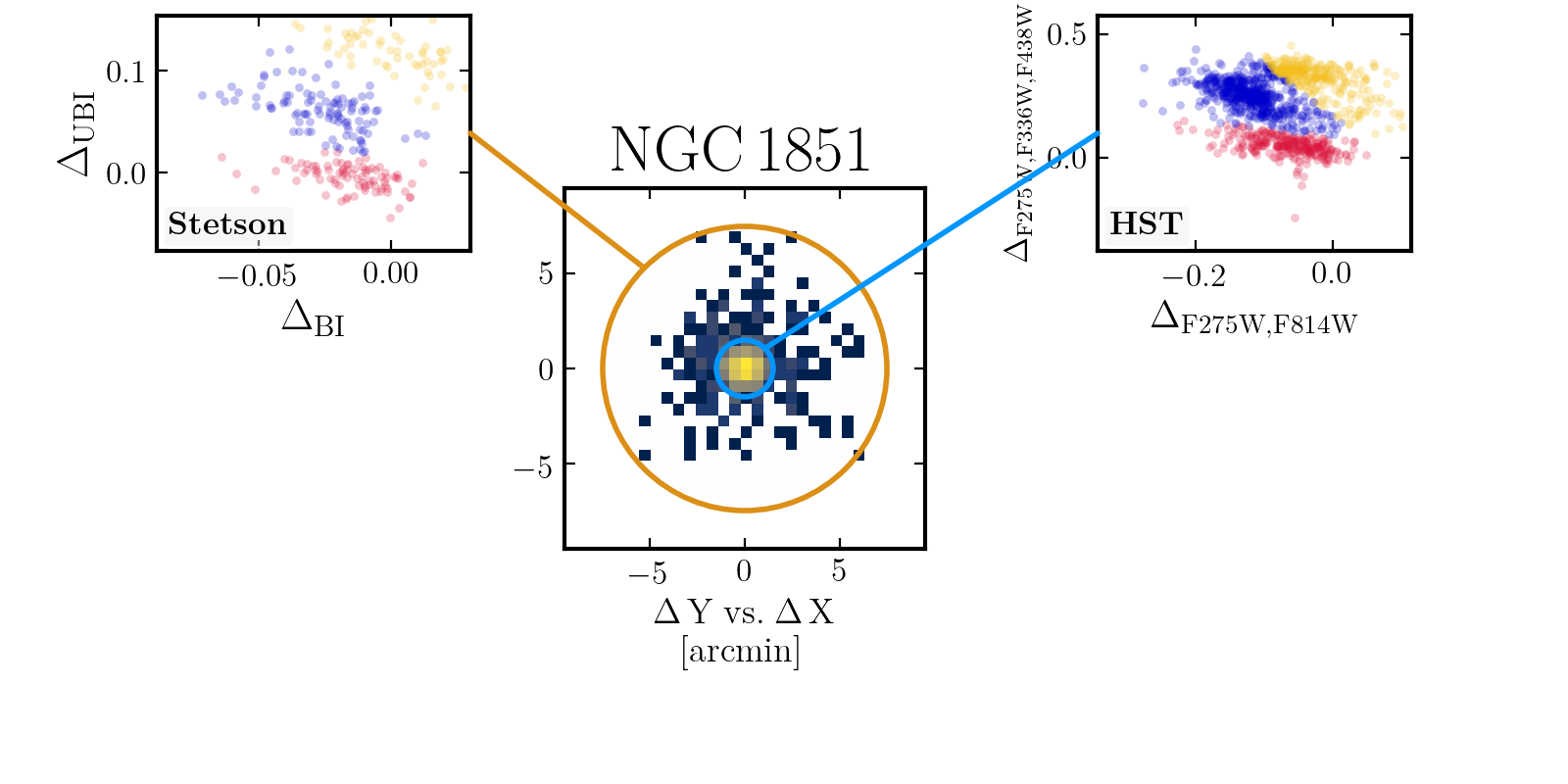}}
    \fbox{\includegraphics[width=0.48\textwidth]{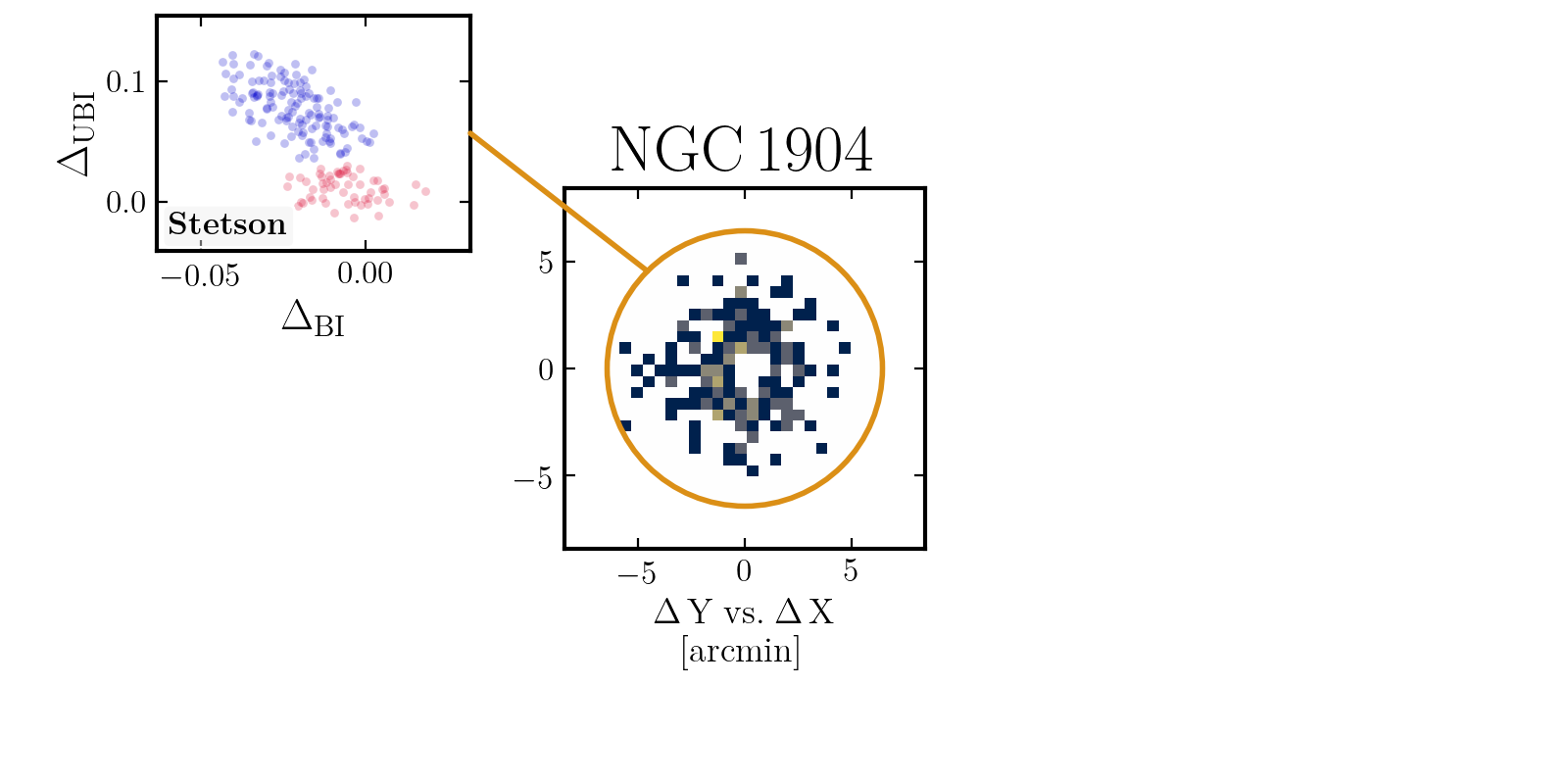}}
    \fbox{\includegraphics[width=0.48\textwidth]{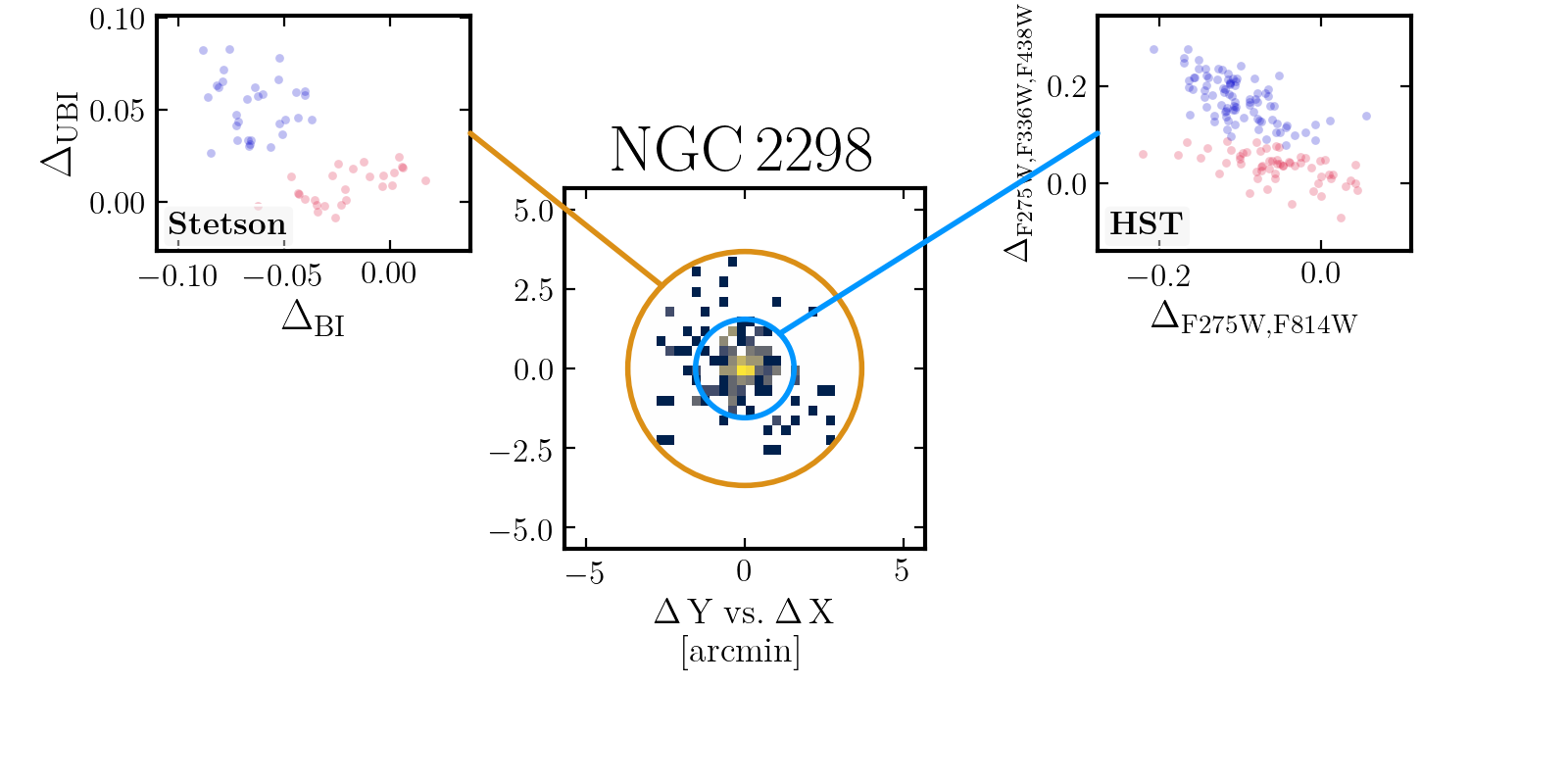}}
    \fbox{\includegraphics[width=0.48\textwidth]{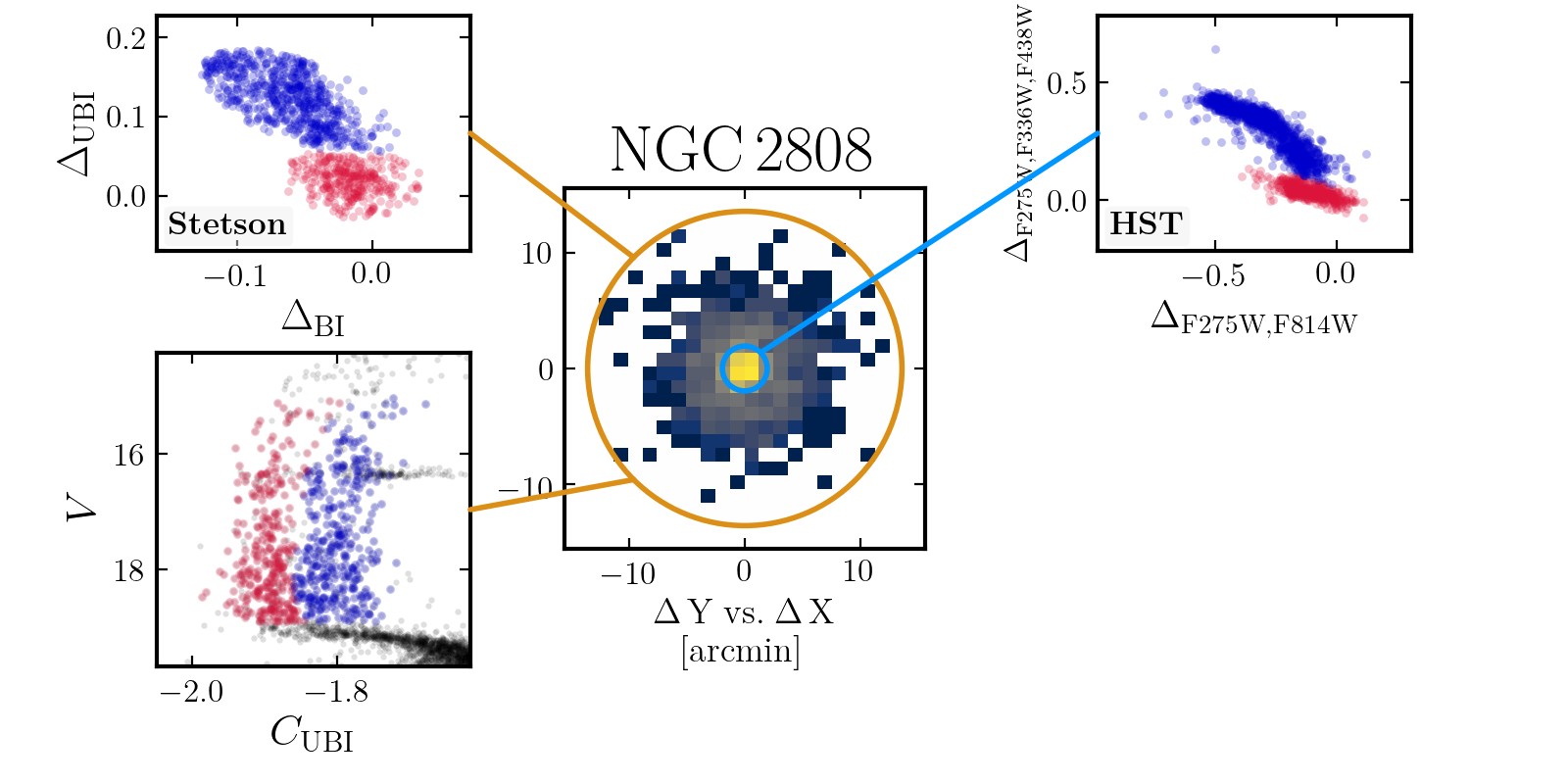}}
    \fbox{\includegraphics[width=0.48\textwidth]{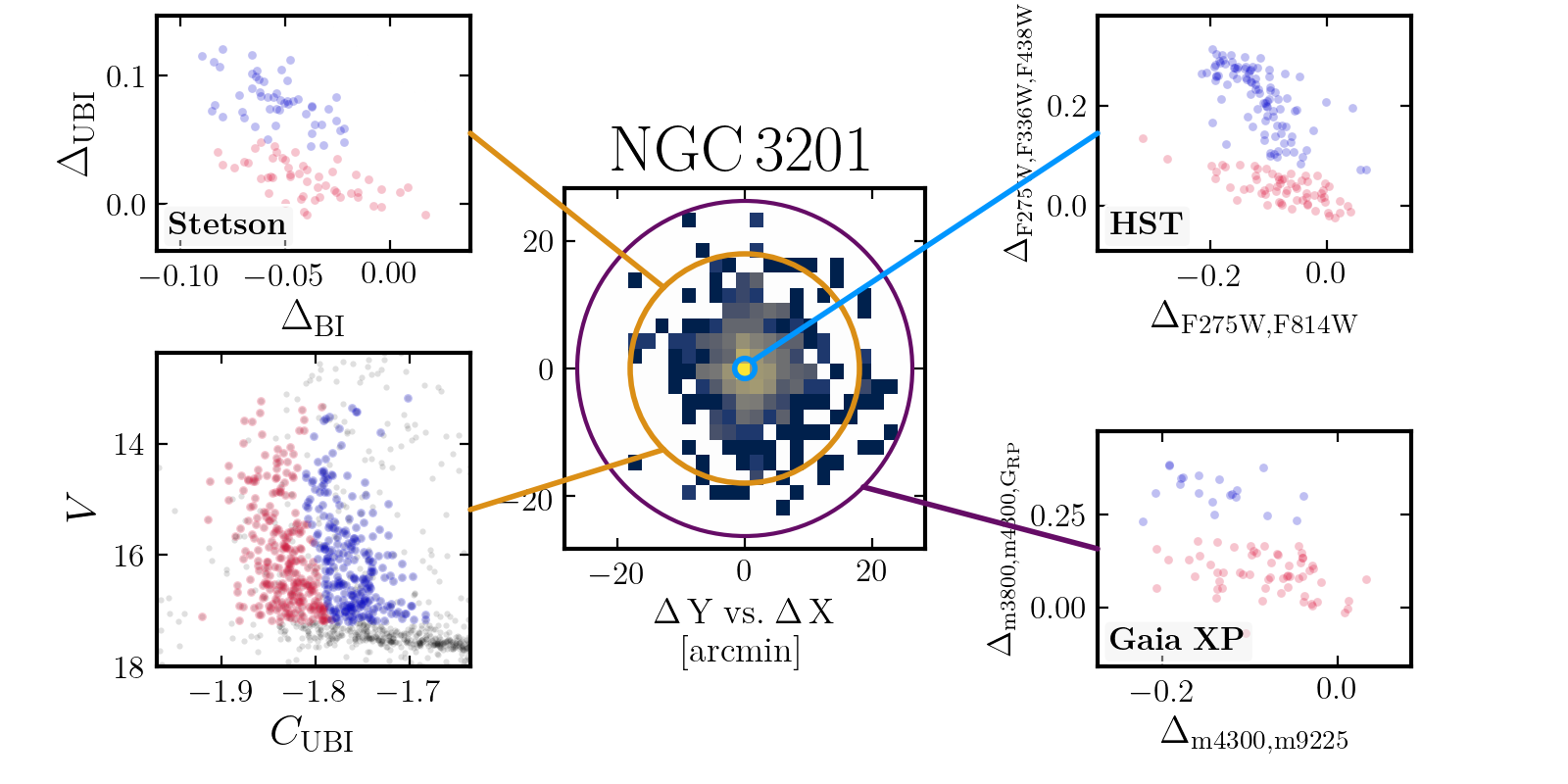}}
    \fbox{\includegraphics[width=0.48\textwidth]{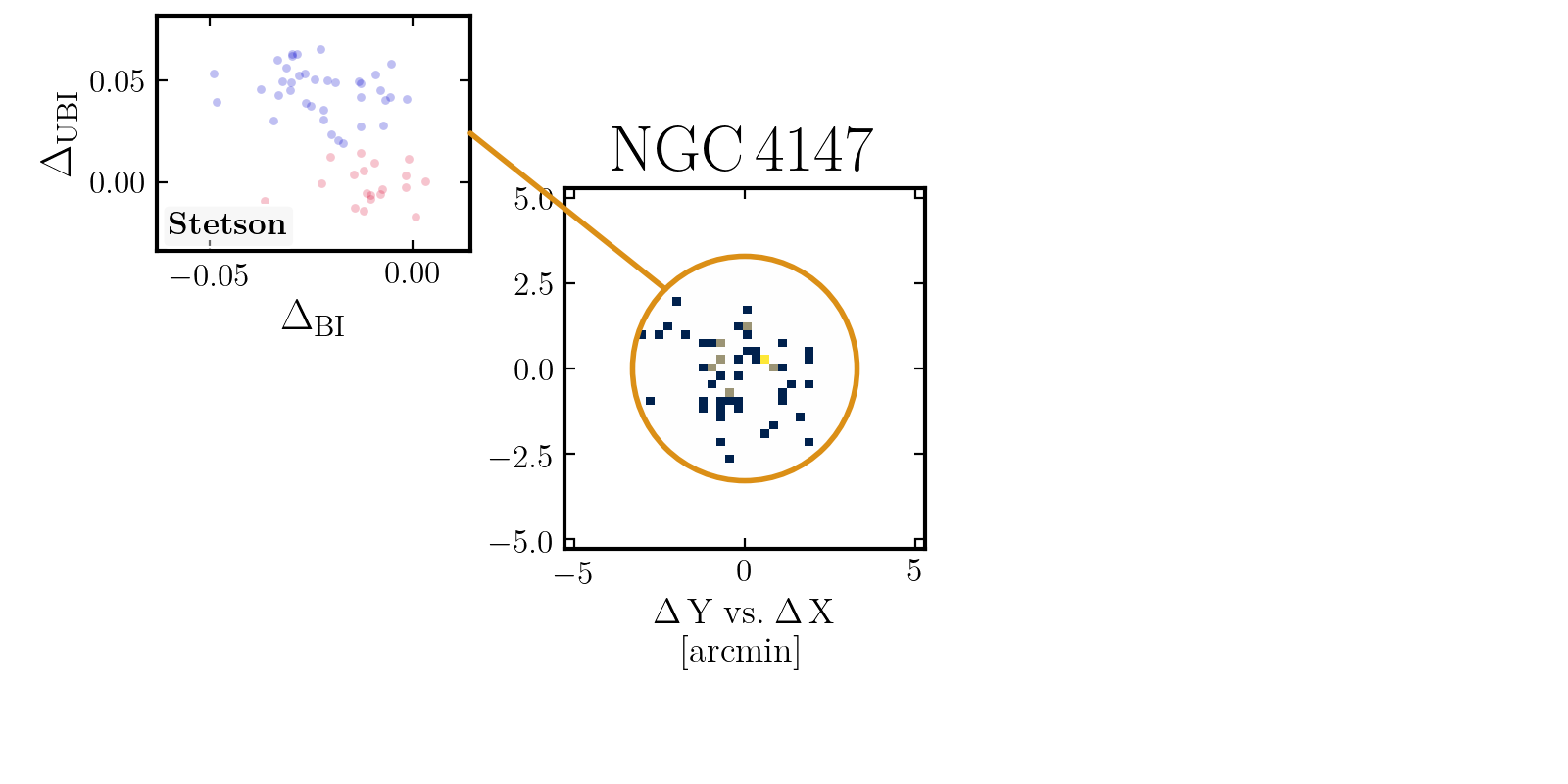}}
    \fbox{\includegraphics[width=0.48\textwidth]{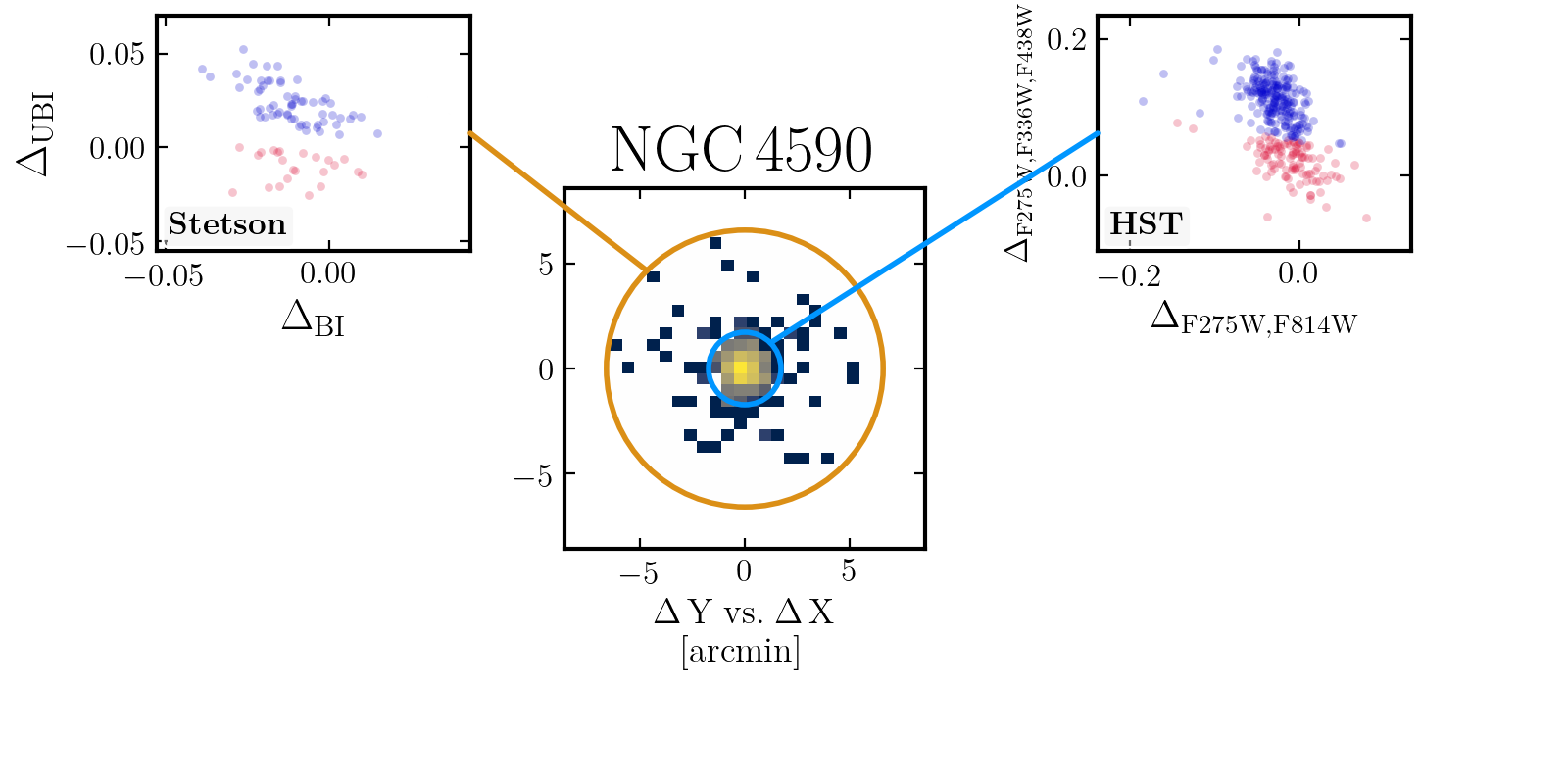}}
    \caption{Photometric diagrams used to identify 1P and 2P stars across the cluster field of view. The left panels display photometric diagrams built with $UBVI$ ground-based photometry from \citet{stetson2019}, while ChMs built with space-based observations are shown on the right side (HST ChMs in the top panel and Gaia XP ChMs in the bottom panel). The central panels indicate the analyzed field of view, with different circles indicating the extension of each dataset. We only show panels whenever a dataset is available. Details of the selection are discussed in Sec.~\ref{sec:data}. 1P, 2P and anomalous stars are shown in red, blue and orange, respectively.}
    \label{fig:mpops1}
\end{figure*}

\begin{figure*}
    \centering
    \fbox{\includegraphics[width=0.48\textwidth]{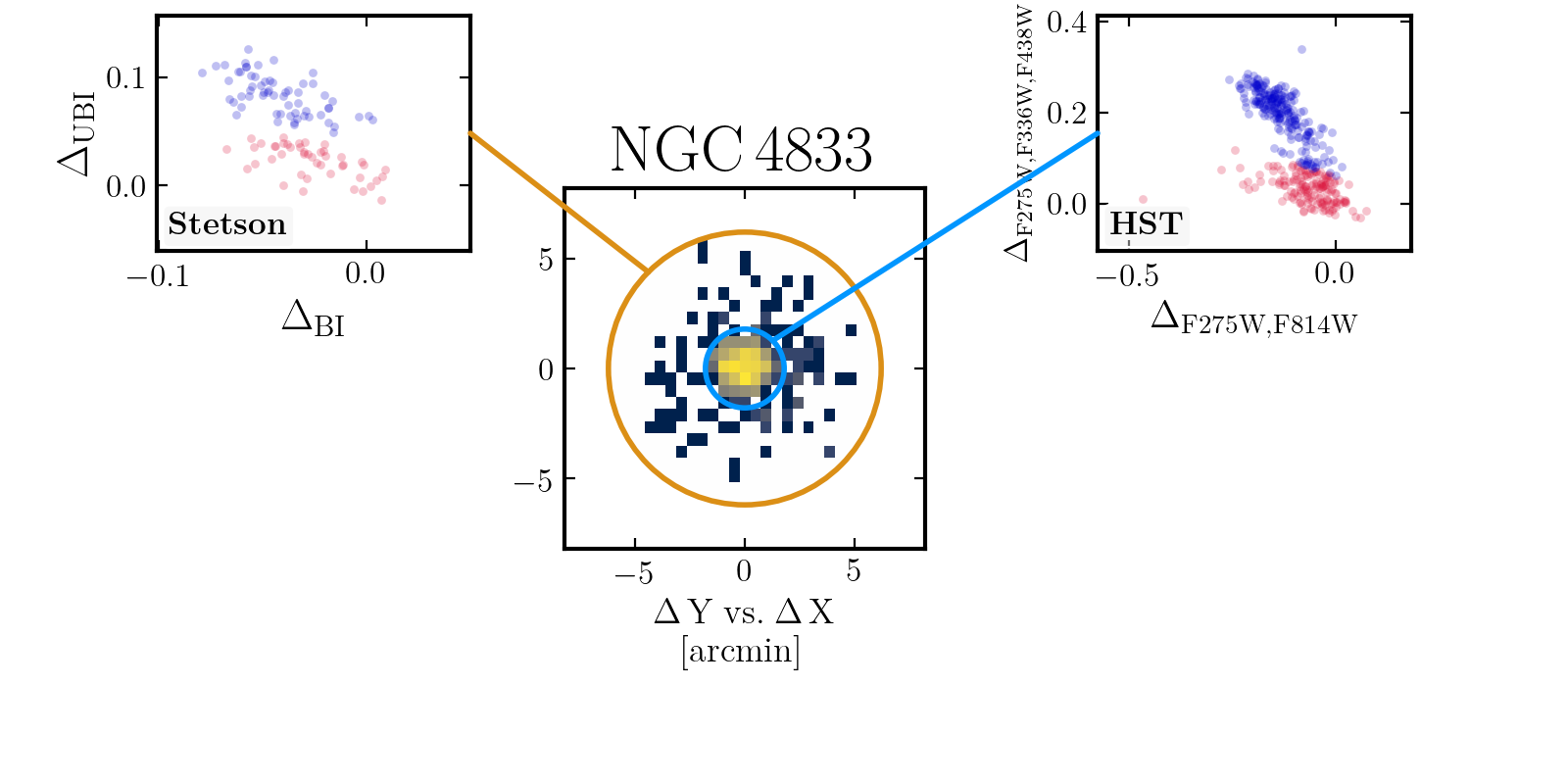}}
    \fbox{\includegraphics[width=0.48\textwidth]{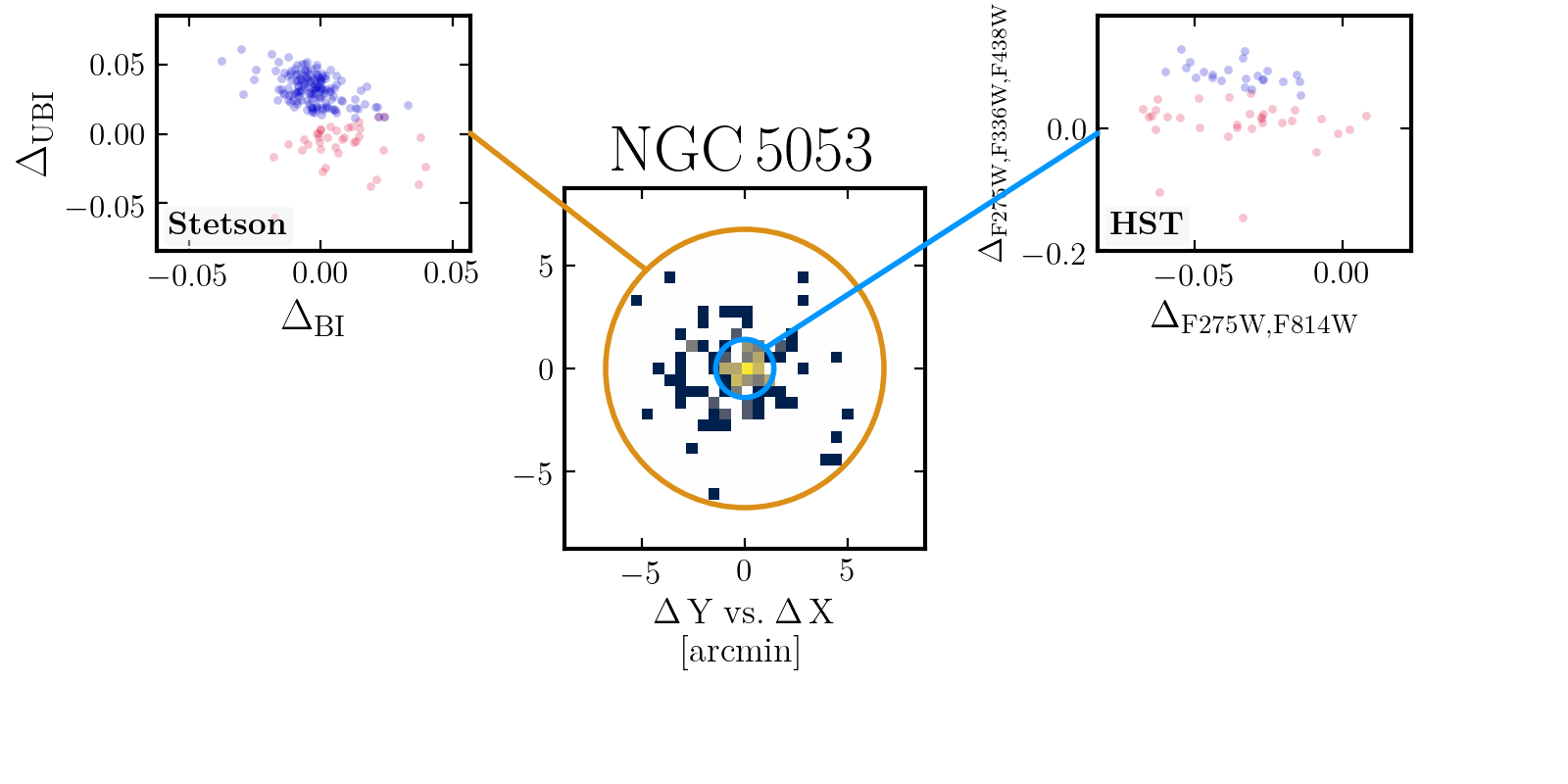}}
    \fbox{\includegraphics[width=0.48\textwidth]{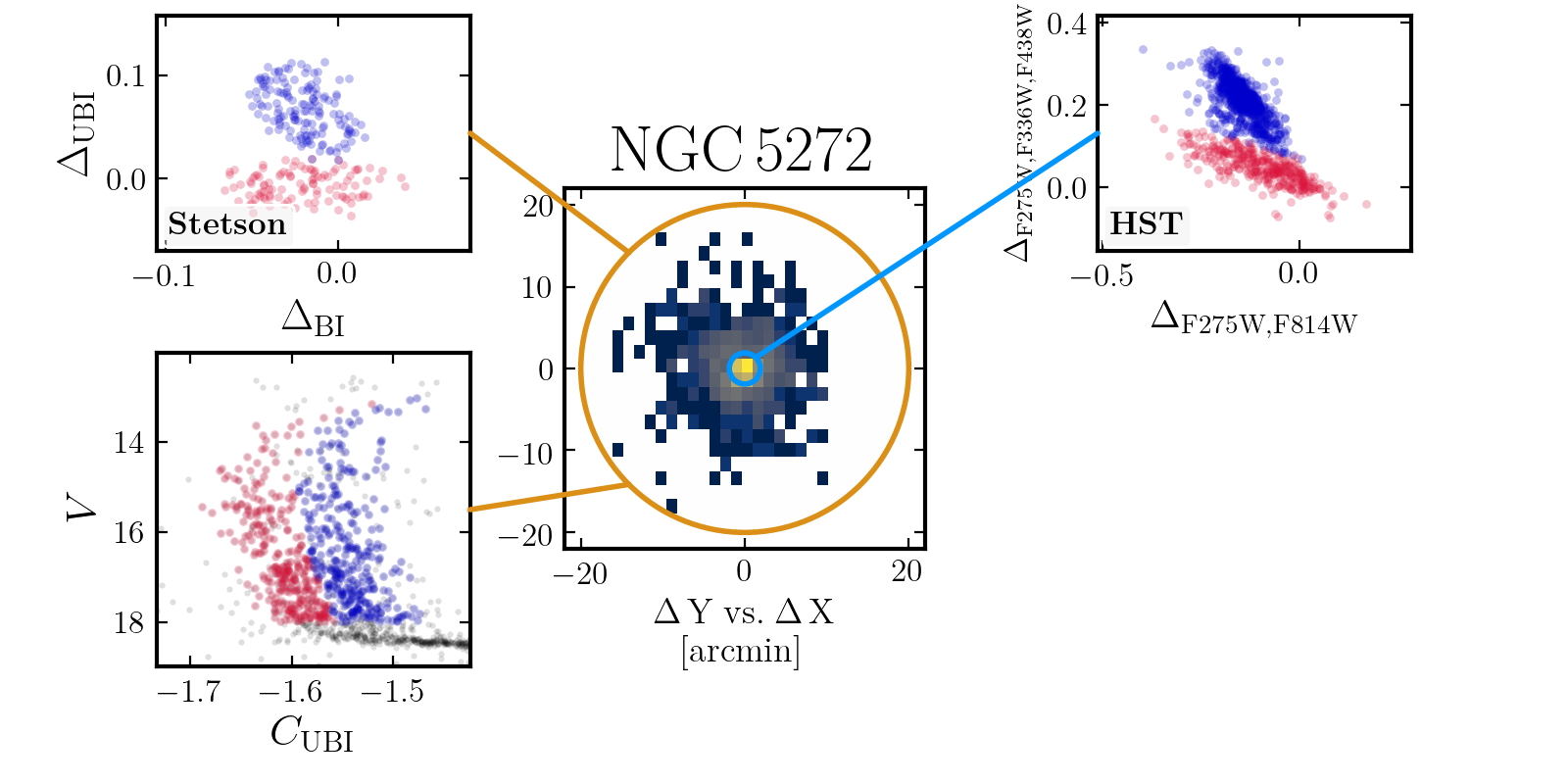}}
    \fbox{\includegraphics[width=0.48\textwidth]{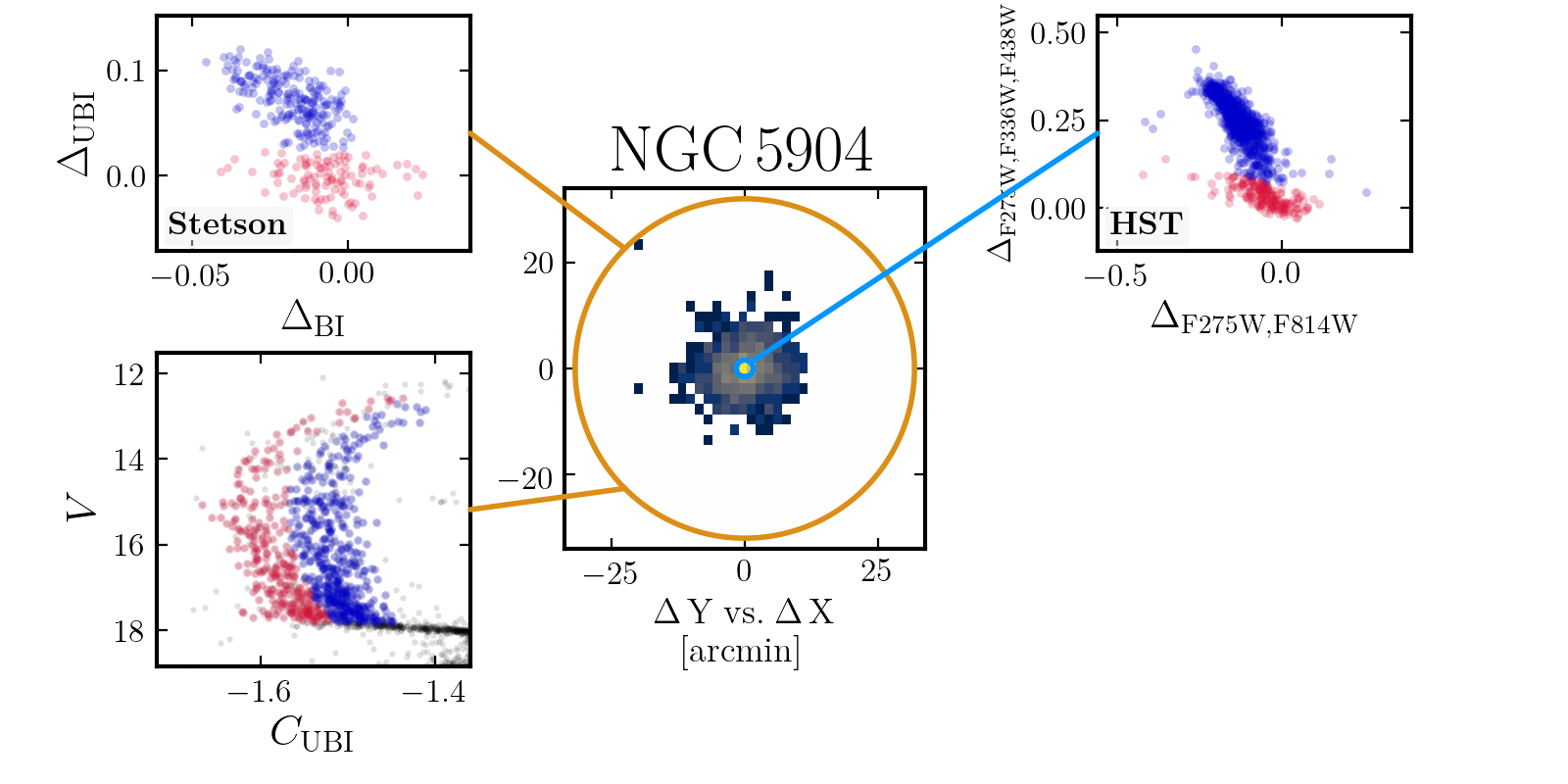}}
    \fbox{\includegraphics[width=0.48\textwidth]{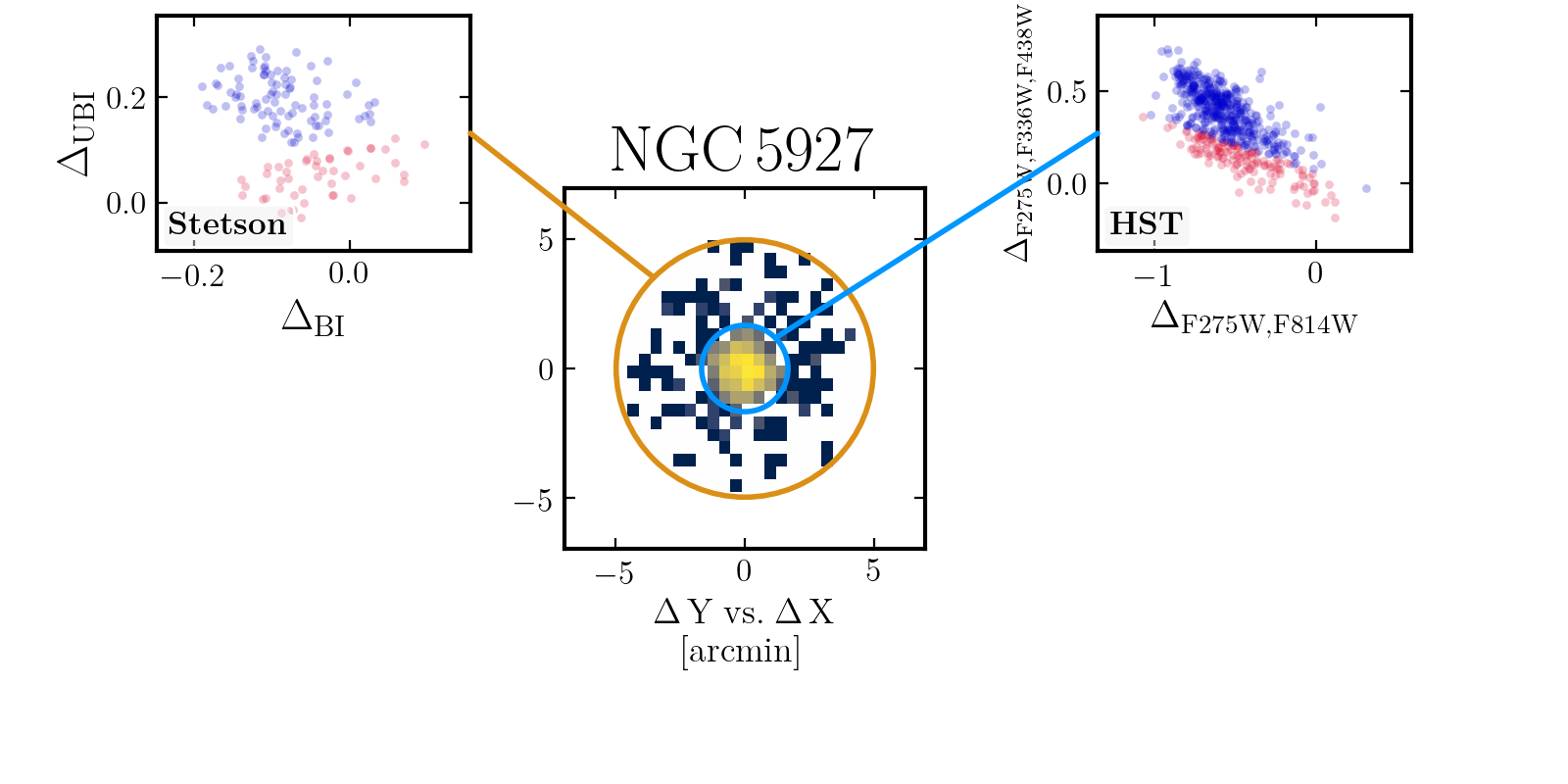}}
    \fbox{\includegraphics[width=0.48\textwidth]{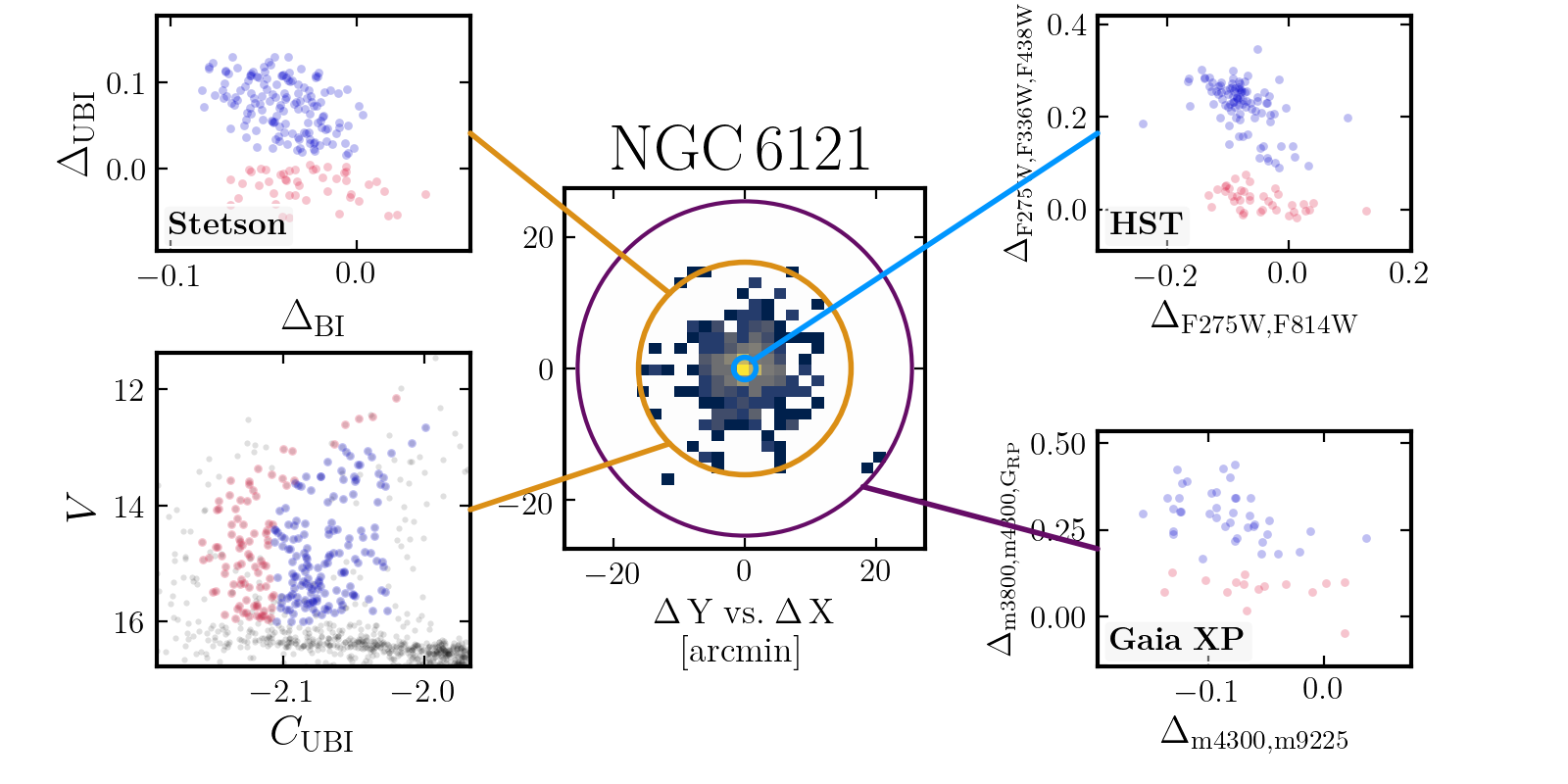}}
    \fbox{\includegraphics[width=0.48\textwidth]{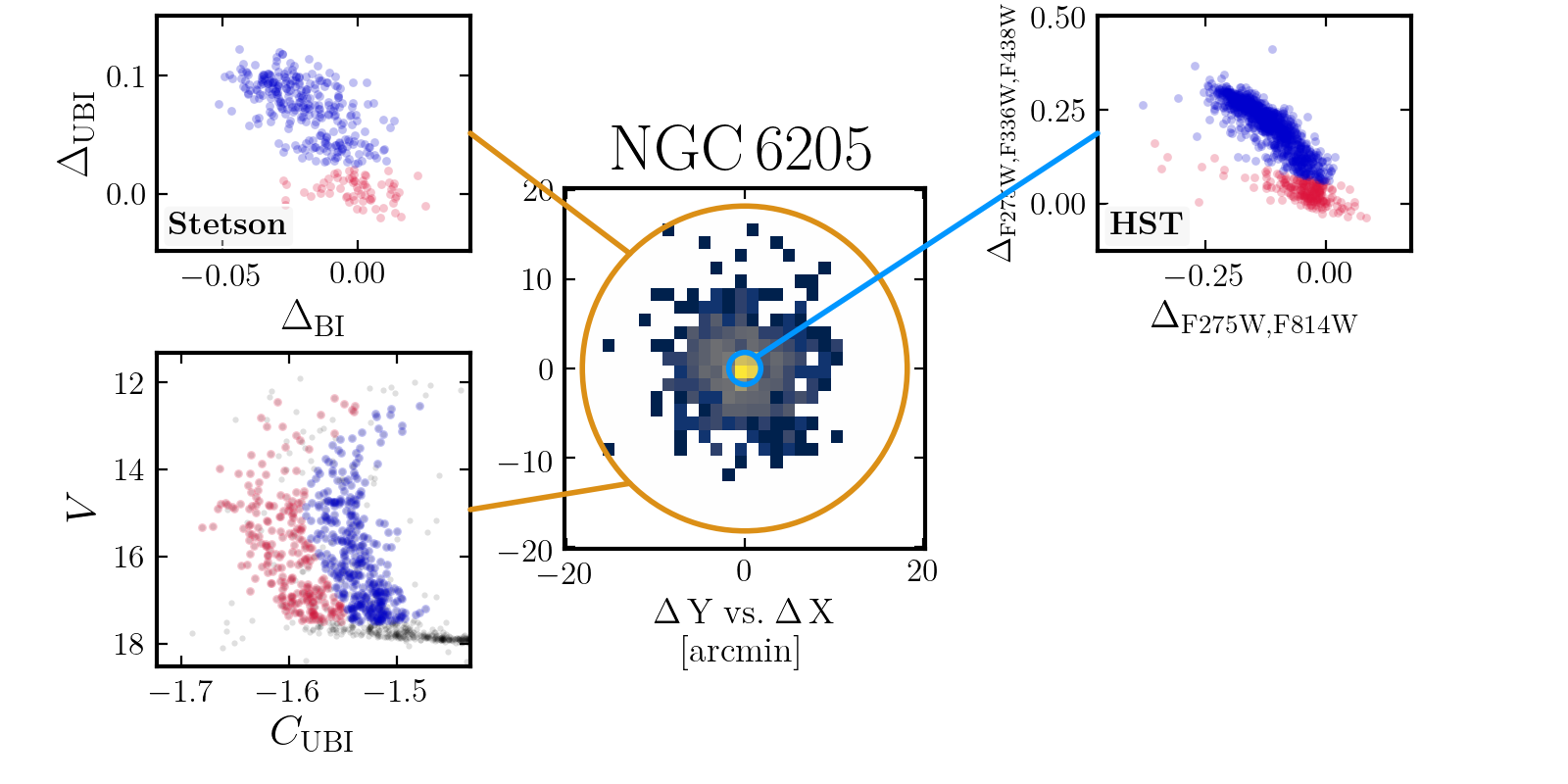}}
    \fbox{\includegraphics[width=0.48\textwidth]{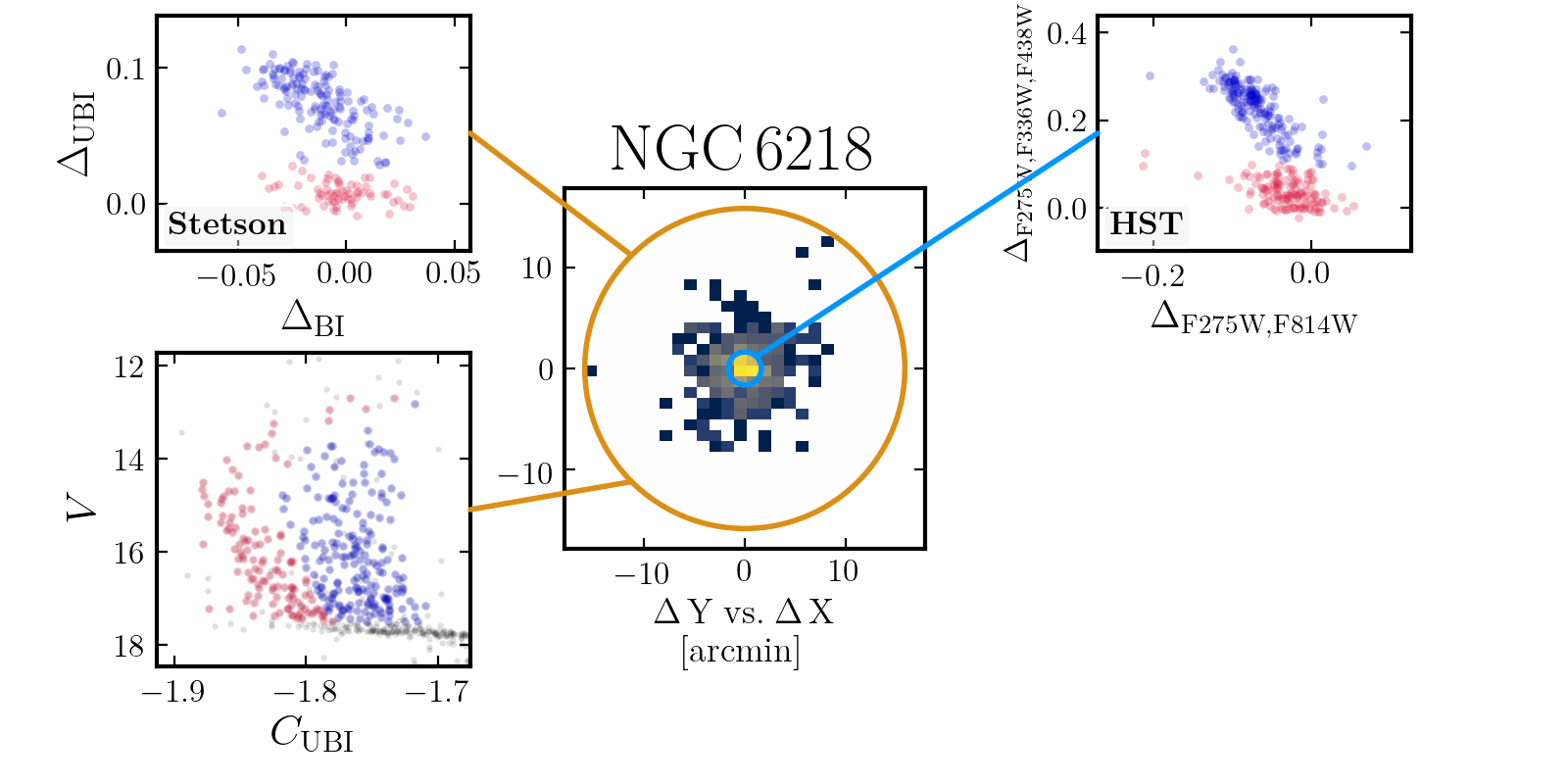}}
    \fbox{\includegraphics[width=0.48\textwidth]{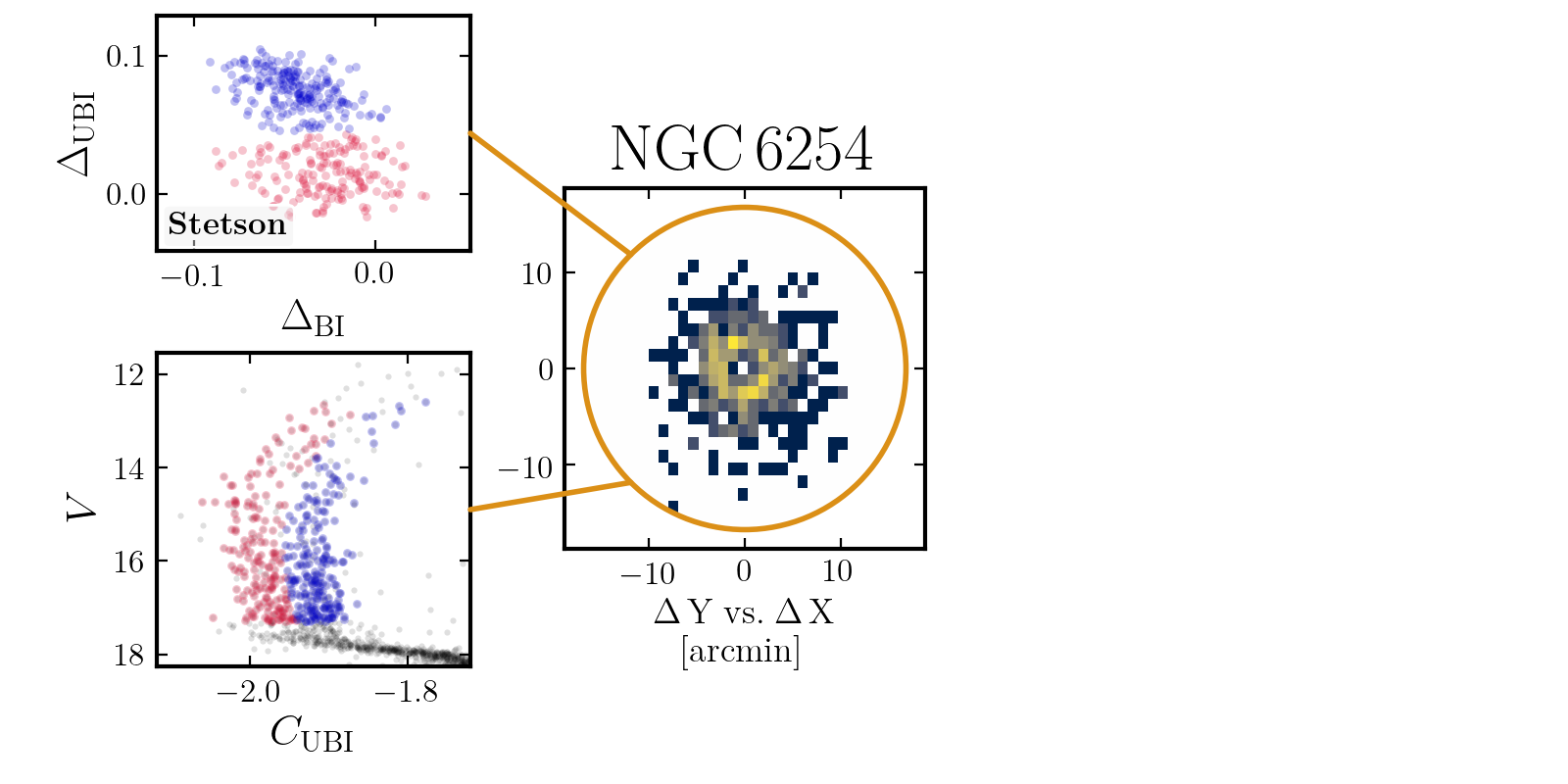}}
    \fbox{\includegraphics[width=0.48\textwidth]{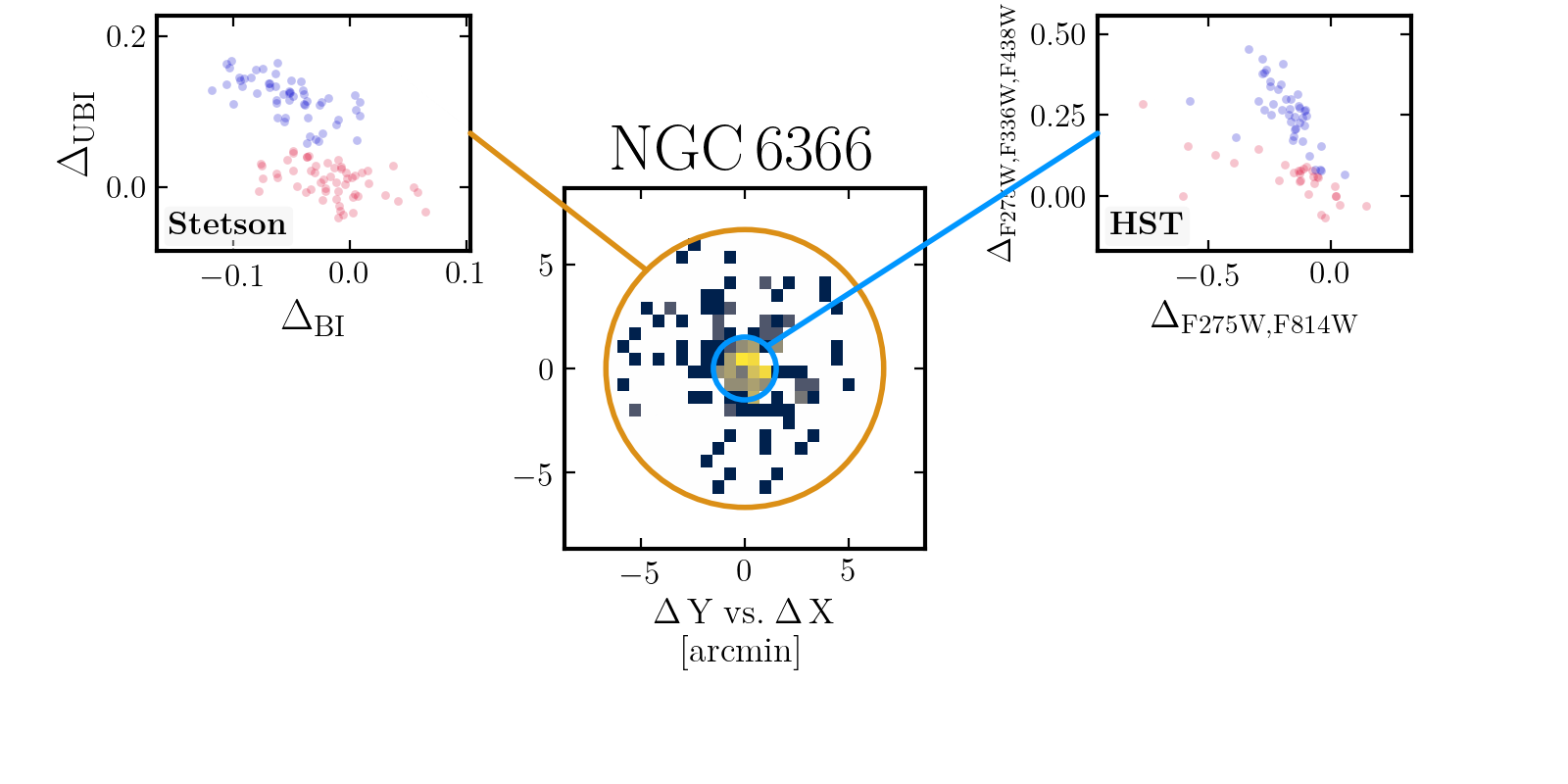}}
    \caption{Same as Fig.~\ref{fig:mpops1}.}
    \label{fig:mpops2}
\end{figure*}

\begin{figure*}
    \centering
    \fbox{\includegraphics[width=0.48\textwidth]{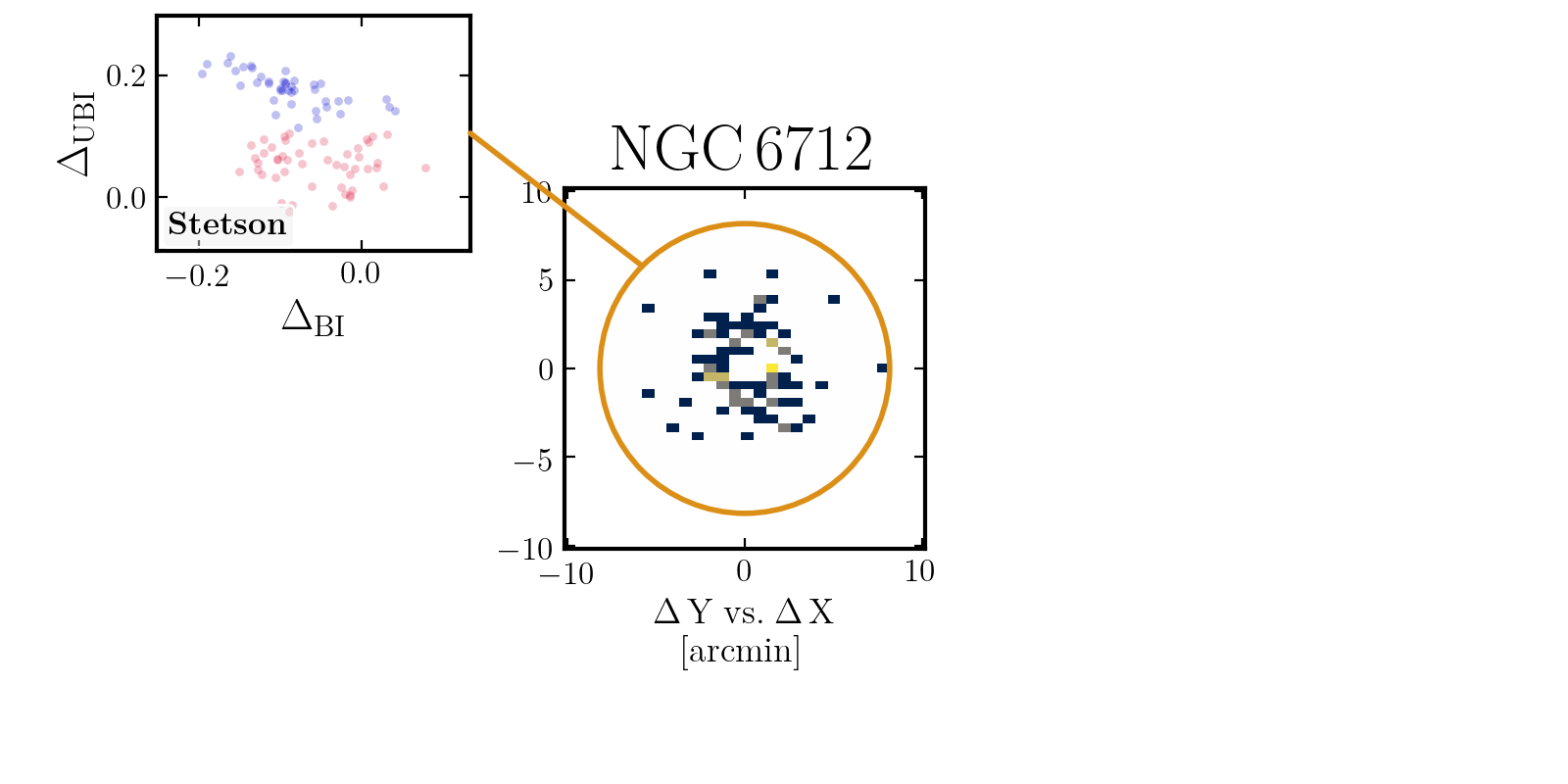}}
    \fbox{\includegraphics[width=0.48\textwidth]{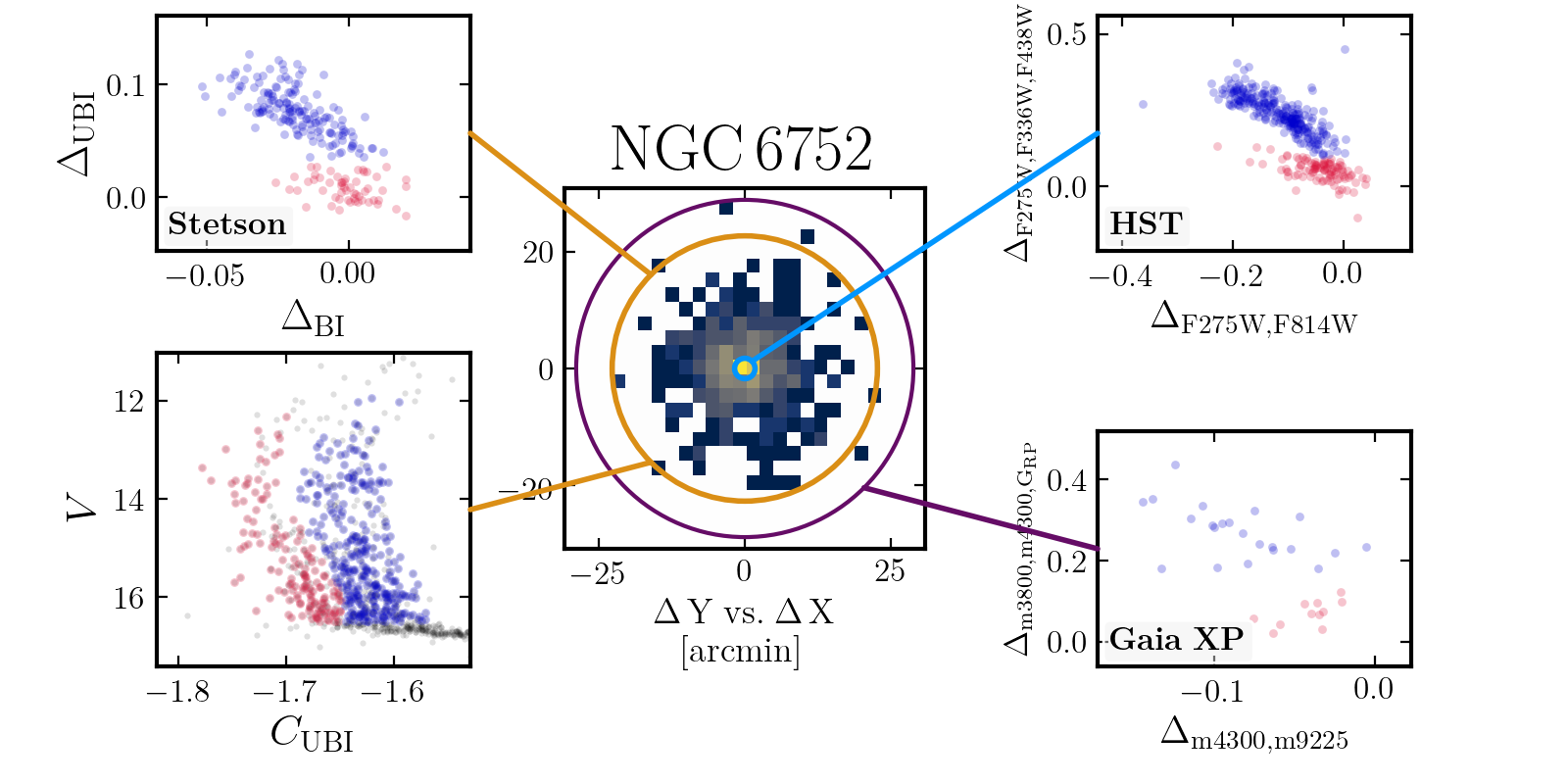}}
    \fbox{\includegraphics[width=0.48\textwidth]{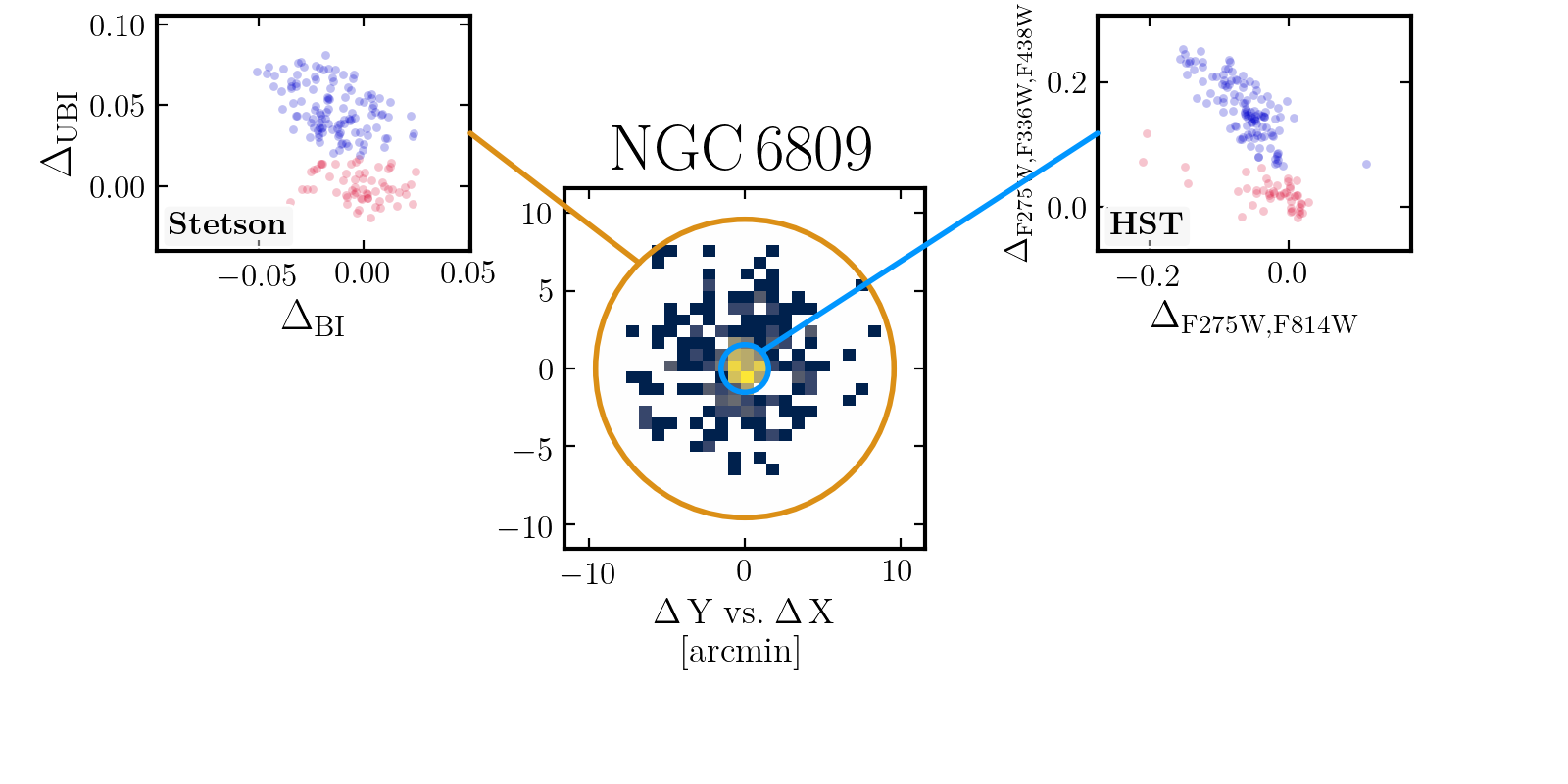}}
    \fbox{\includegraphics[width=0.48\textwidth]{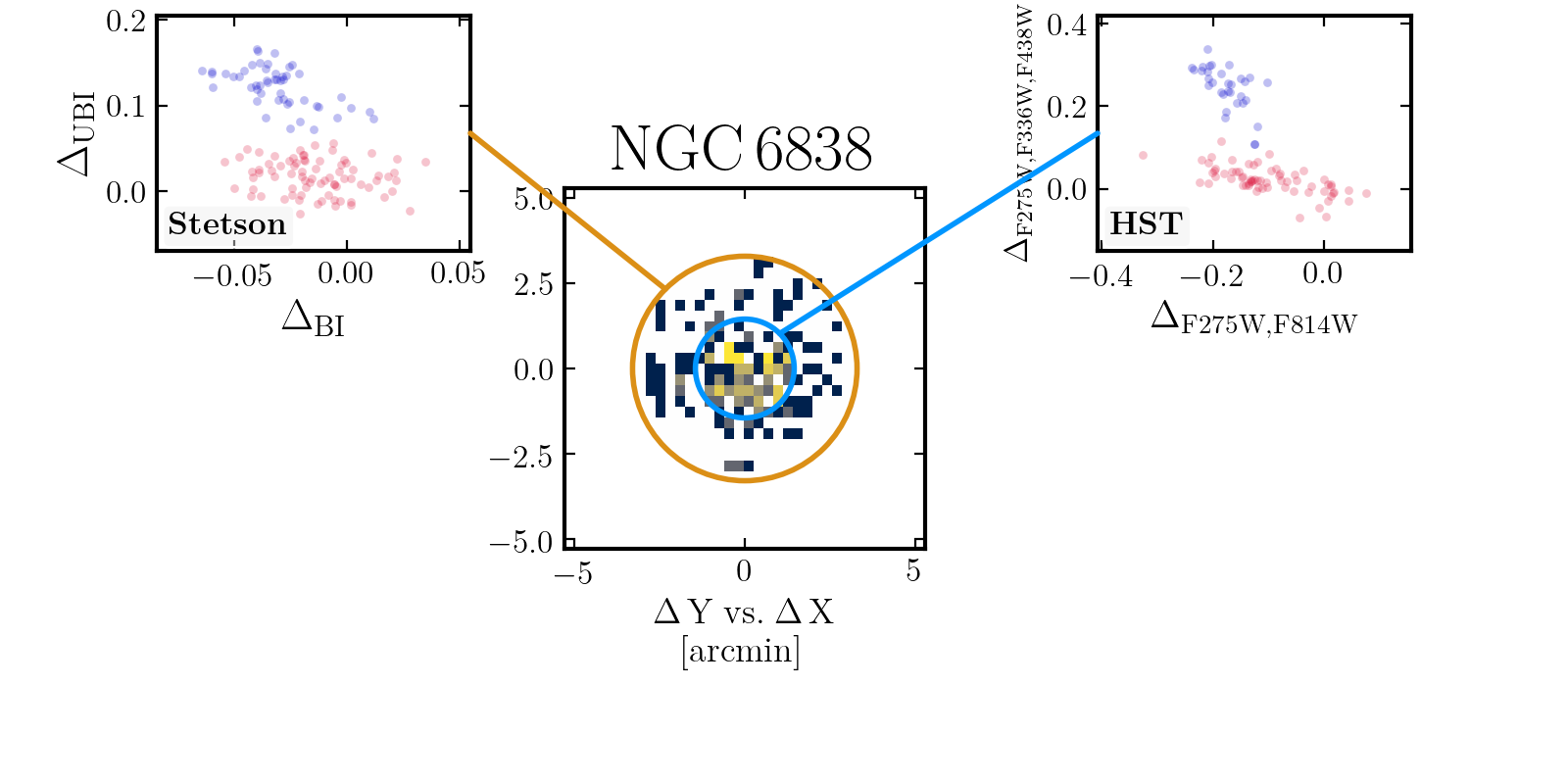}}
    \fbox{\includegraphics[width=0.48\textwidth]{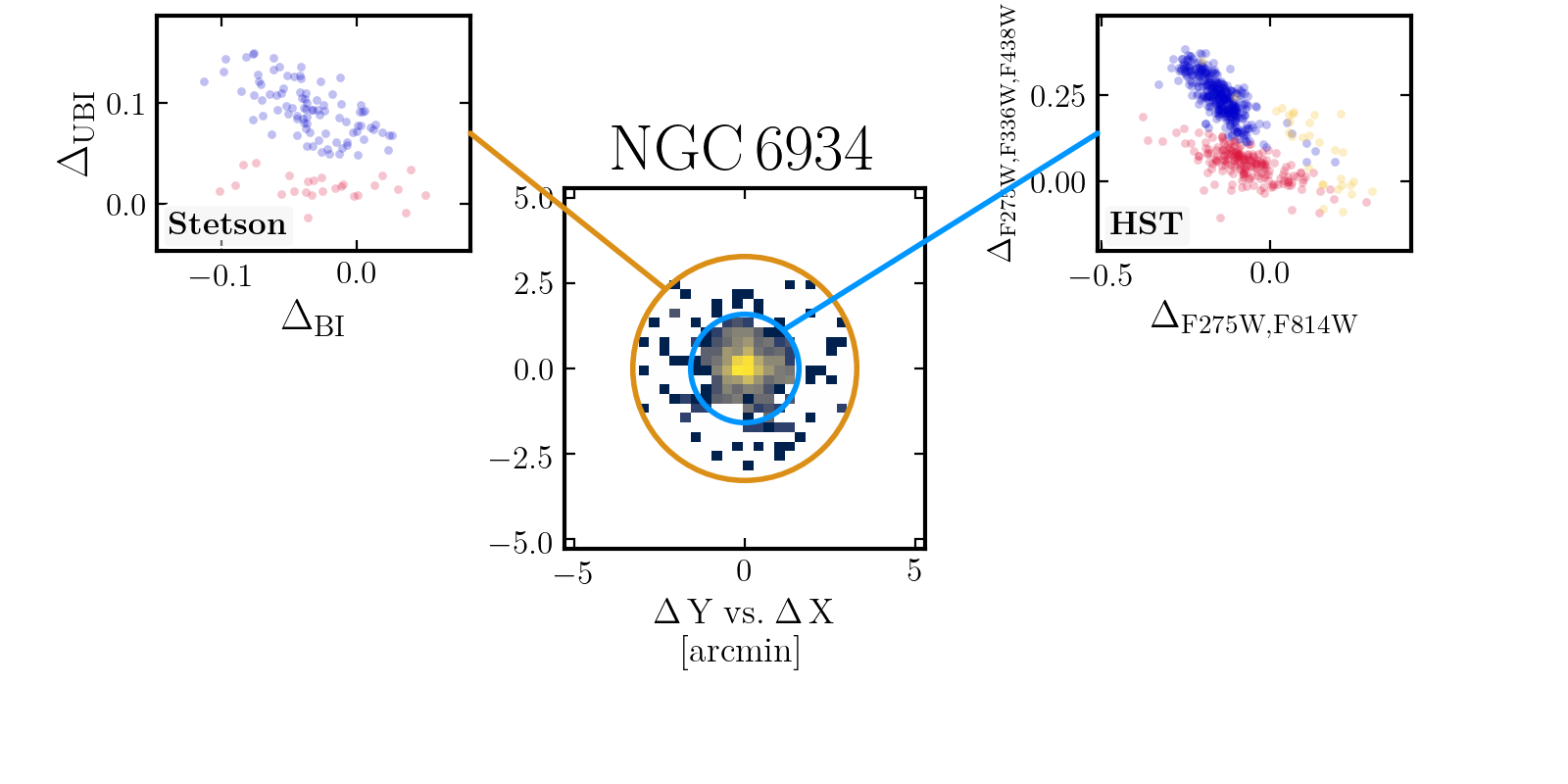}}
    \fbox{\includegraphics[width=0.48\textwidth]{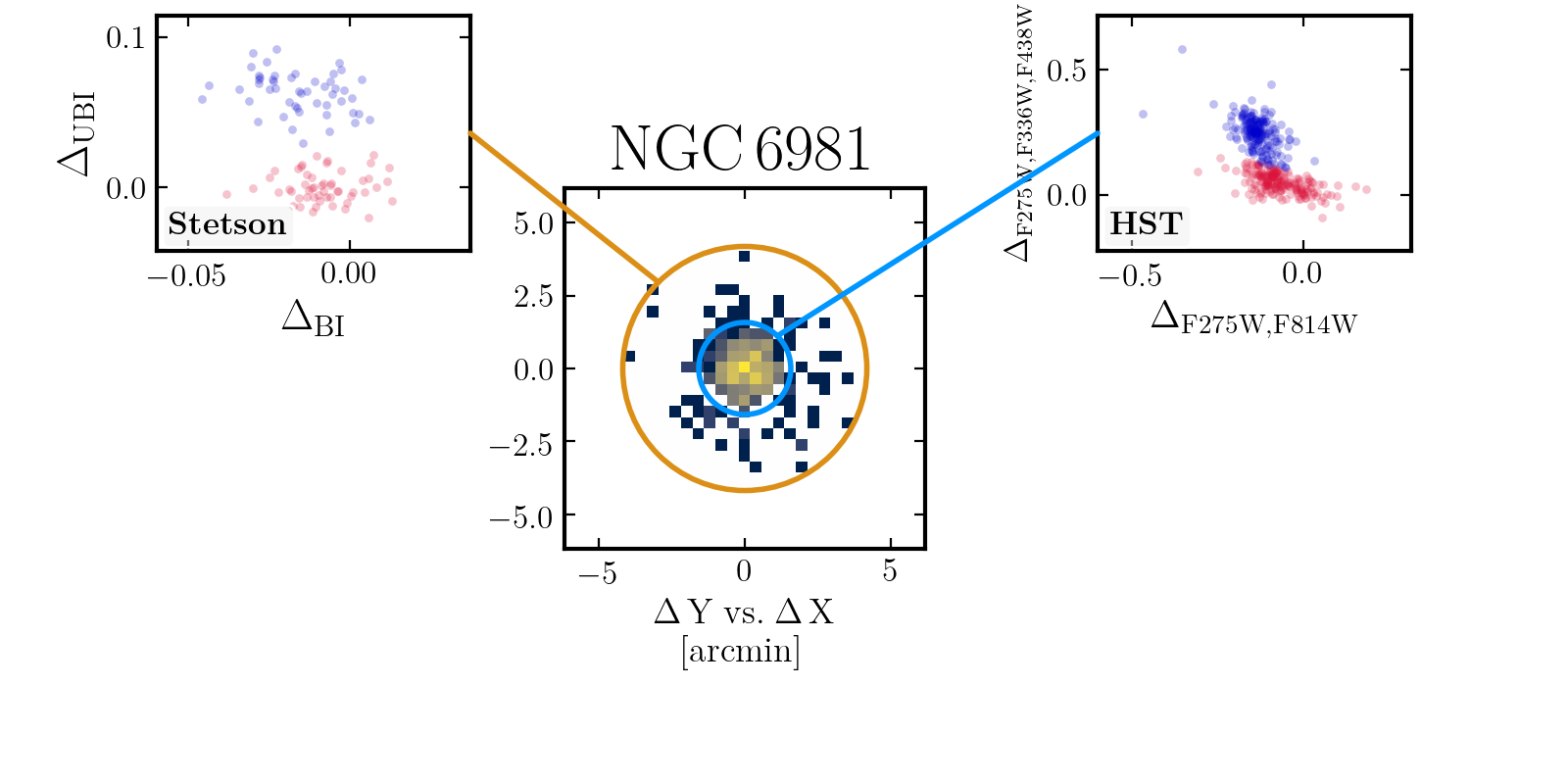}}
    \fbox{\includegraphics[width=0.48\textwidth]{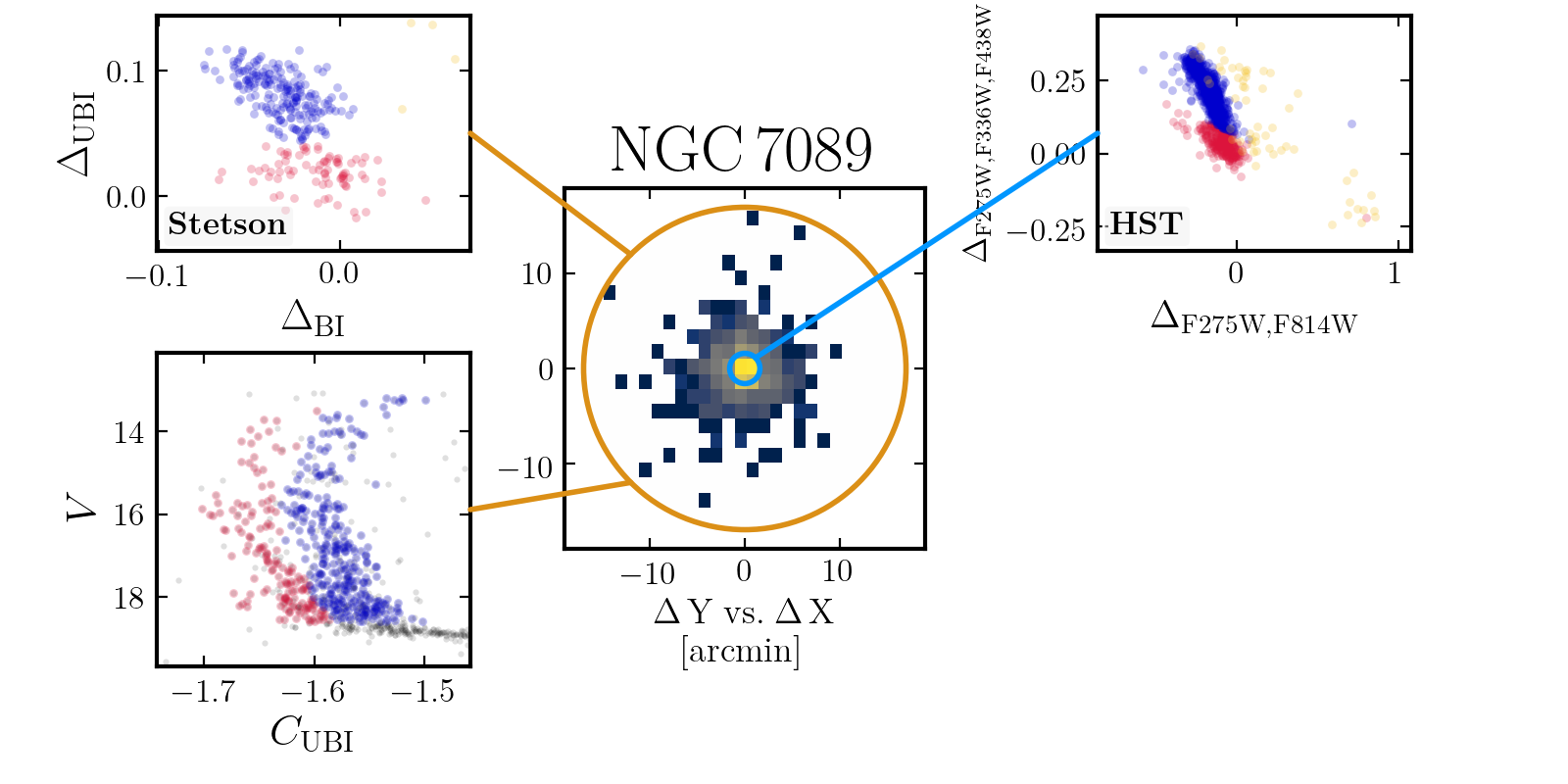}}
    \fbox{\includegraphics[width=0.48\textwidth]{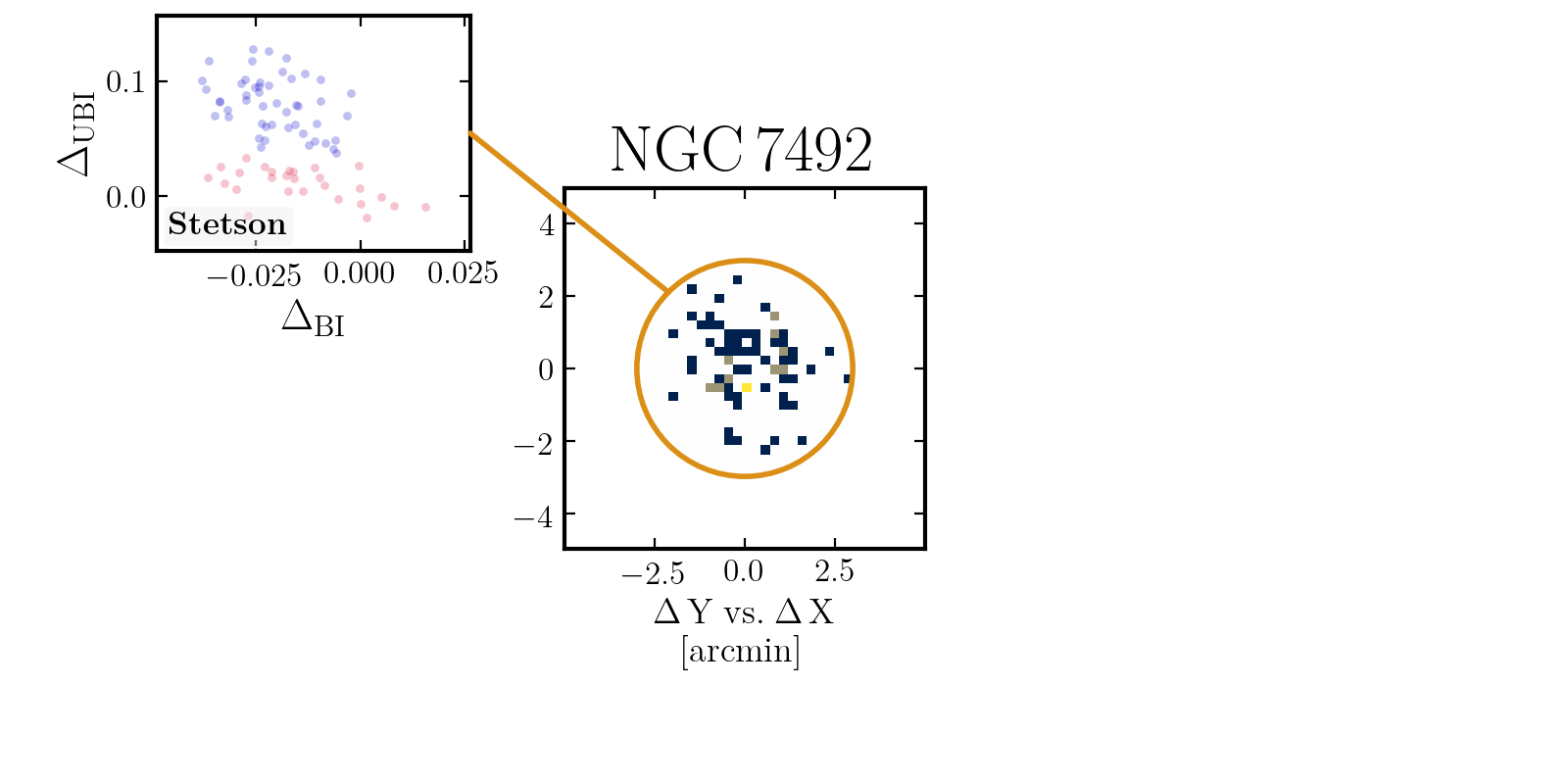}}
    \caption{Same as Fig.~\ref{fig:mpops1}.}
    \label{fig:mpops3}
\end{figure*}

In this Appendix we show the selection of MPs for all the 28 analyzed clusters. As discussed in Sec.~\ref{sec:data}, we used a combination of HST and ground-based photometry \citep{milone2017, nardiello2018, stetson2019, jang2022}, and Gaia XP synthetic photometry \citep{mehta2024}. In Fig.~\ref{fig:mpops1}-~\ref{fig:mpops3} we display the analyzed field of view in the central panel, together with the photometric diagrams used to separate 1P and 2P, always shown with red and blue colors respectively. Specifically, ground-based photometry is shown in the leftmost panels, with the ChMs from \citet{jang2022} shown in the top one. HST and Gaia XP ChM are instead displayed in the right panels, respectively in the top and bottom one. The coloured circles in the central panels indicate the analyzed regions. 

Concerning ground-based observations, we first used the selection in the ChM whenever available, and we adopted the $\cubi$ selection for stars without ChM information. MPs outside the orange circles are selected by means of Gaia XP synthethic photometry. We remind here that Gaia XP synthethic photometry is only available for four clusters, namely NGC\,0104, NGC\,3201, NGC\,6121 and NGC\,6752 \citep[see][for a complete discussion]{mehta2024}. For \typeii clusters in our sample, we show anomalous stars as orange points in HST and ground-based ChMs.

\section{Individual profiles of all clusters}
\label{app:individual profiles}
\begin{figure*}
    \centering
    \includegraphics[width=0.90\textwidth]{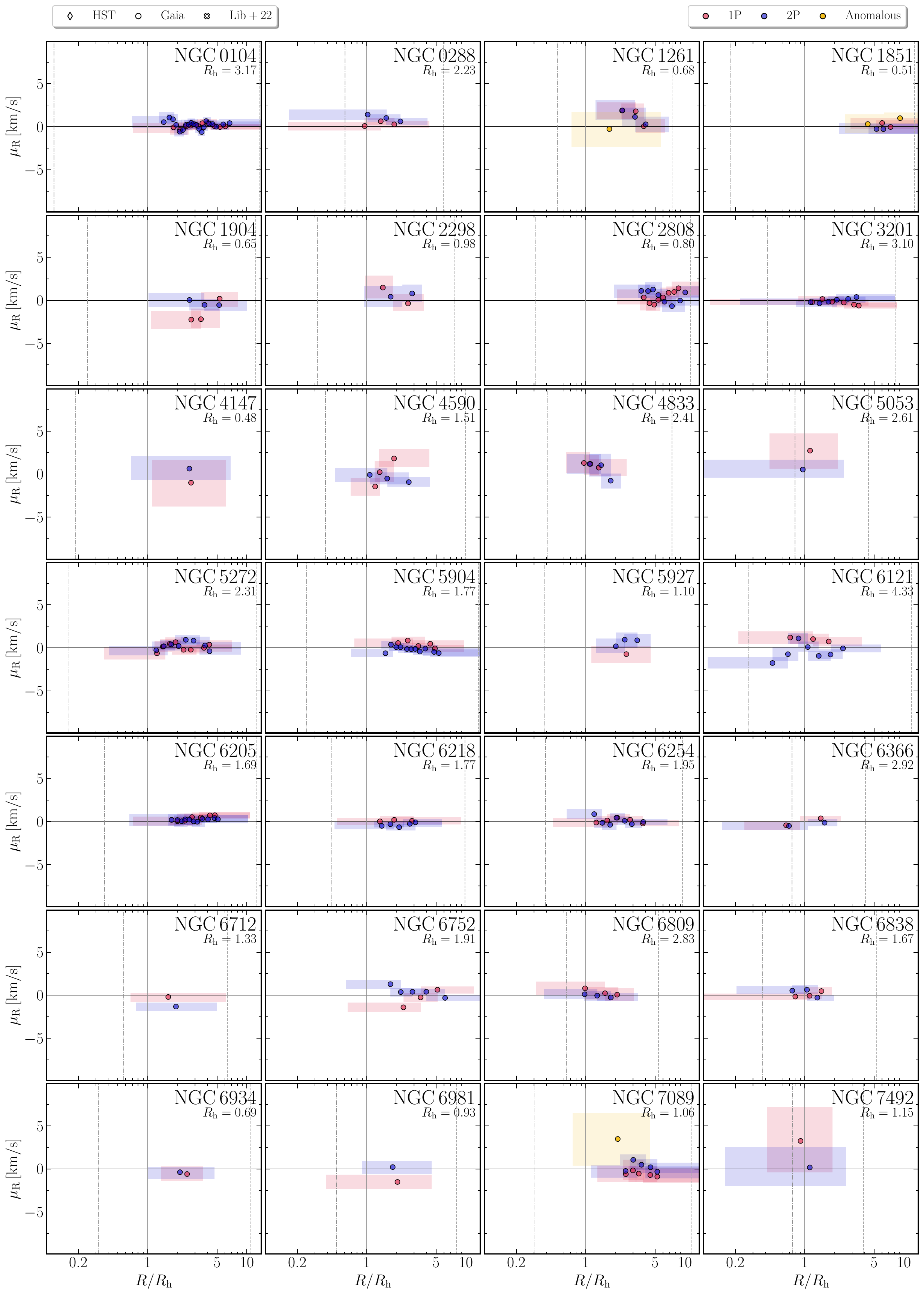}
    \caption{Mean motion of the radial component, determined as discussed in the Sec.~\ref{sec:internal dynamics}. The radial coordinate is normalized to the half-light radius, indicated in the inset in units of arcmin, while dynamical quantities are shown in $\mathrm{km/s}$. Red, blue and orange colours represent the profiles of 1P, 2P and anomalous stars (if present).}
    \label{fig:radmed}
\end{figure*}

\begin{figure*}
    \centering
    \includegraphics[width=0.90\textwidth]{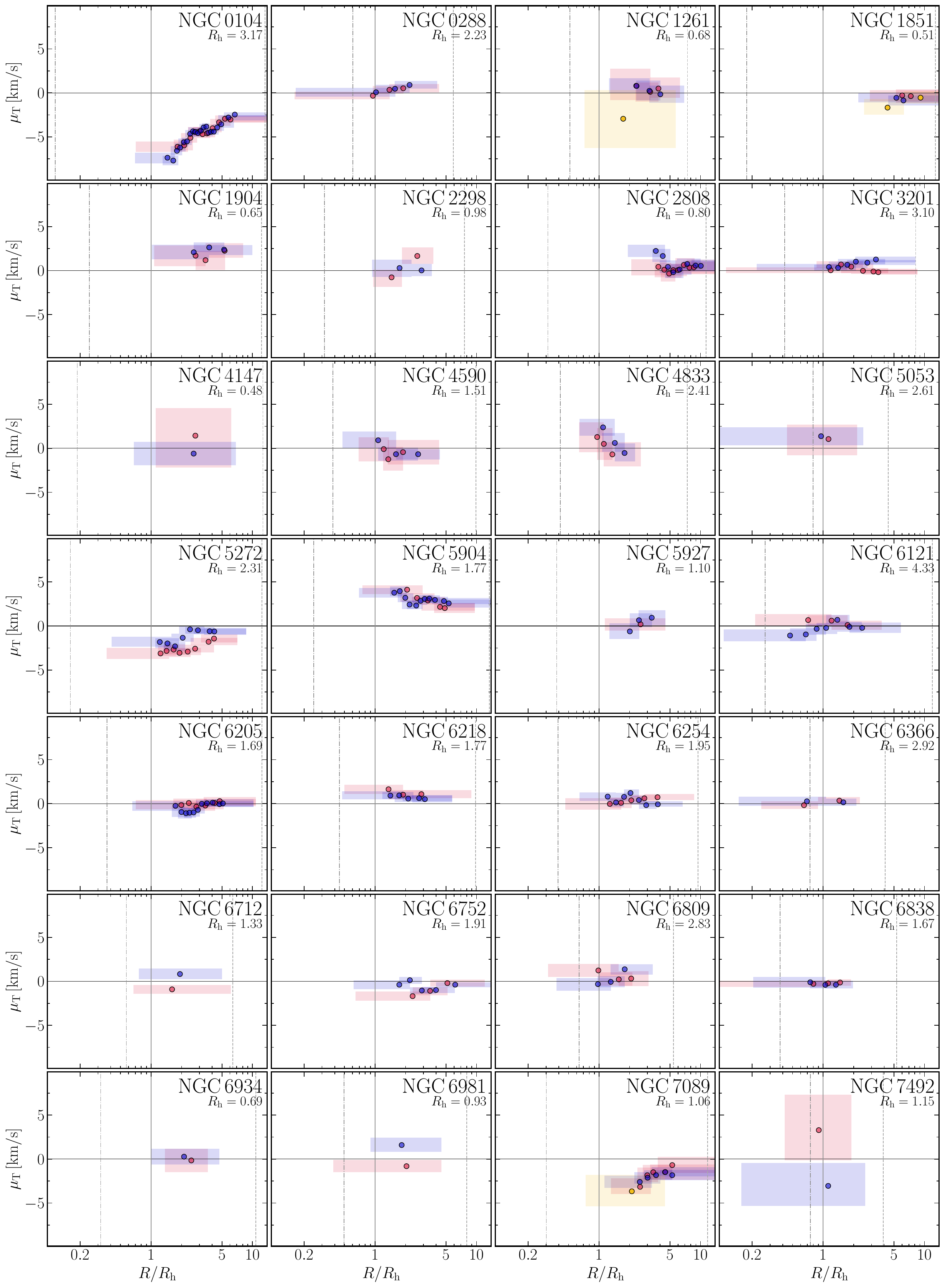}
    \caption{Same as Fig.~\ref{fig:radmed} for the tangential component.}
    \label{fig:tanmed}
\end{figure*}

\begin{figure*}
    \centering
    \includegraphics[width=0.90\textwidth]{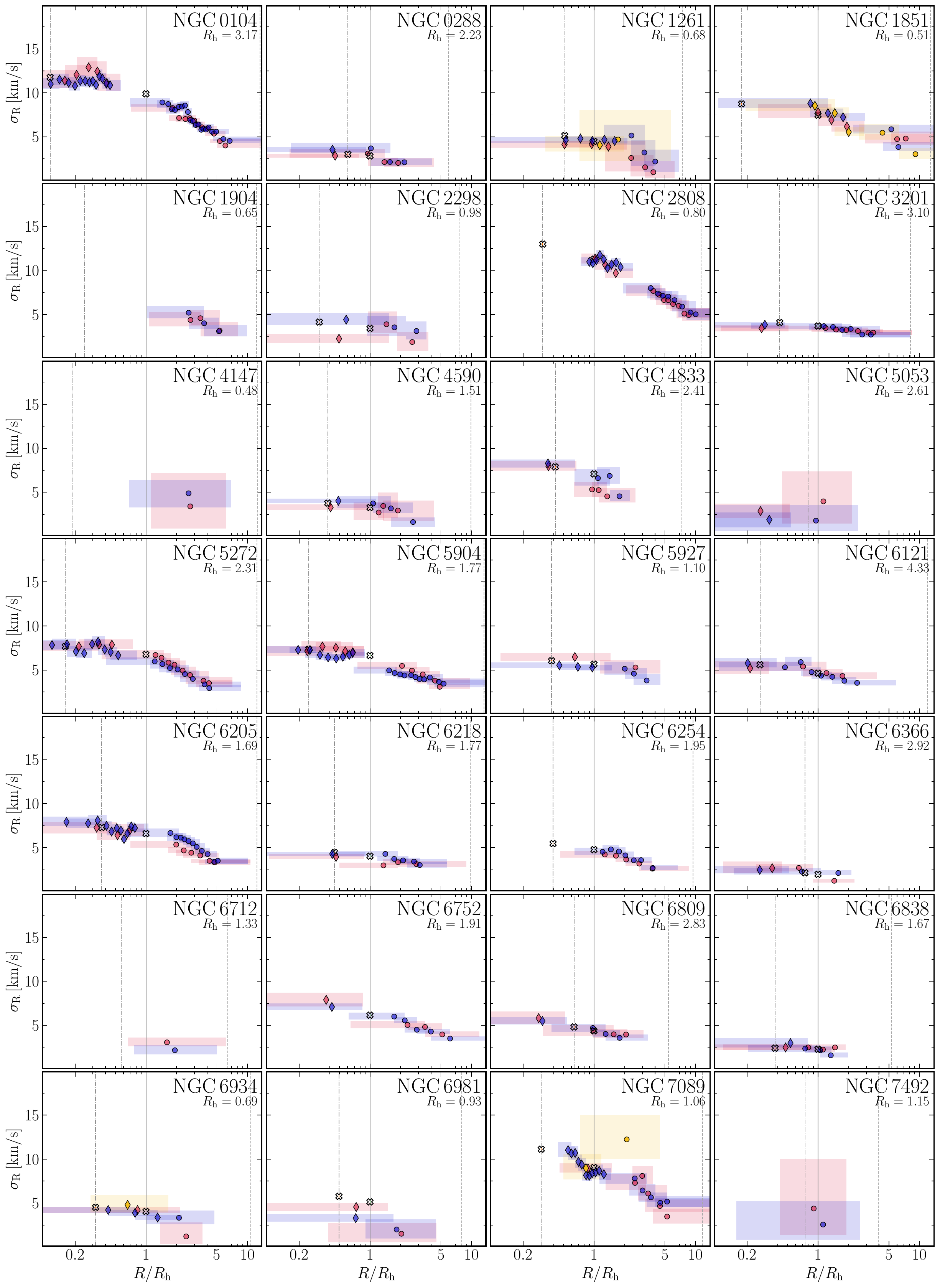}
    \caption{Dispersion profile along the radial direction. The values of the 2-dimensional dispersion computed in the cluster center and at the core and half-light radii in \citet{libralato2022} are indicated by white thick crosses, while the colour-coding is as in Fig.~\ref{fig:radmed}. Results inferred from HST and Gaia proper motions are shown with diamonds and circles, respectively.}
    \label{fig:raddis}
\end{figure*}

\begin{figure*}
    \centering
    \includegraphics[width=0.90\textwidth]{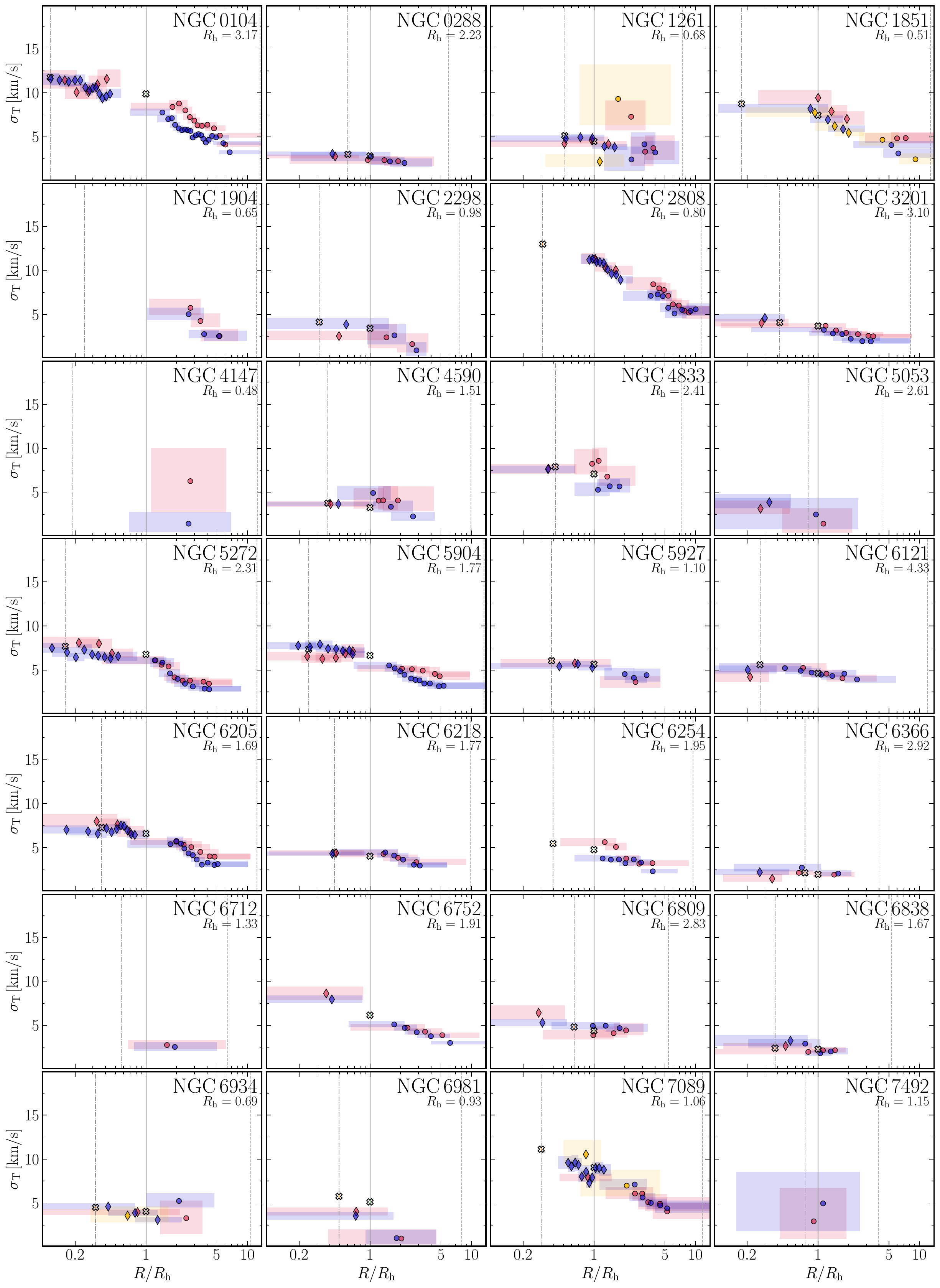}
    \caption{Same as Fig.~\ref{fig:raddis} for the tangential component.}
    \label{fig:tandis}
\end{figure*}

\begin{figure*}
    \centering
    \includegraphics[width=0.90\textwidth]{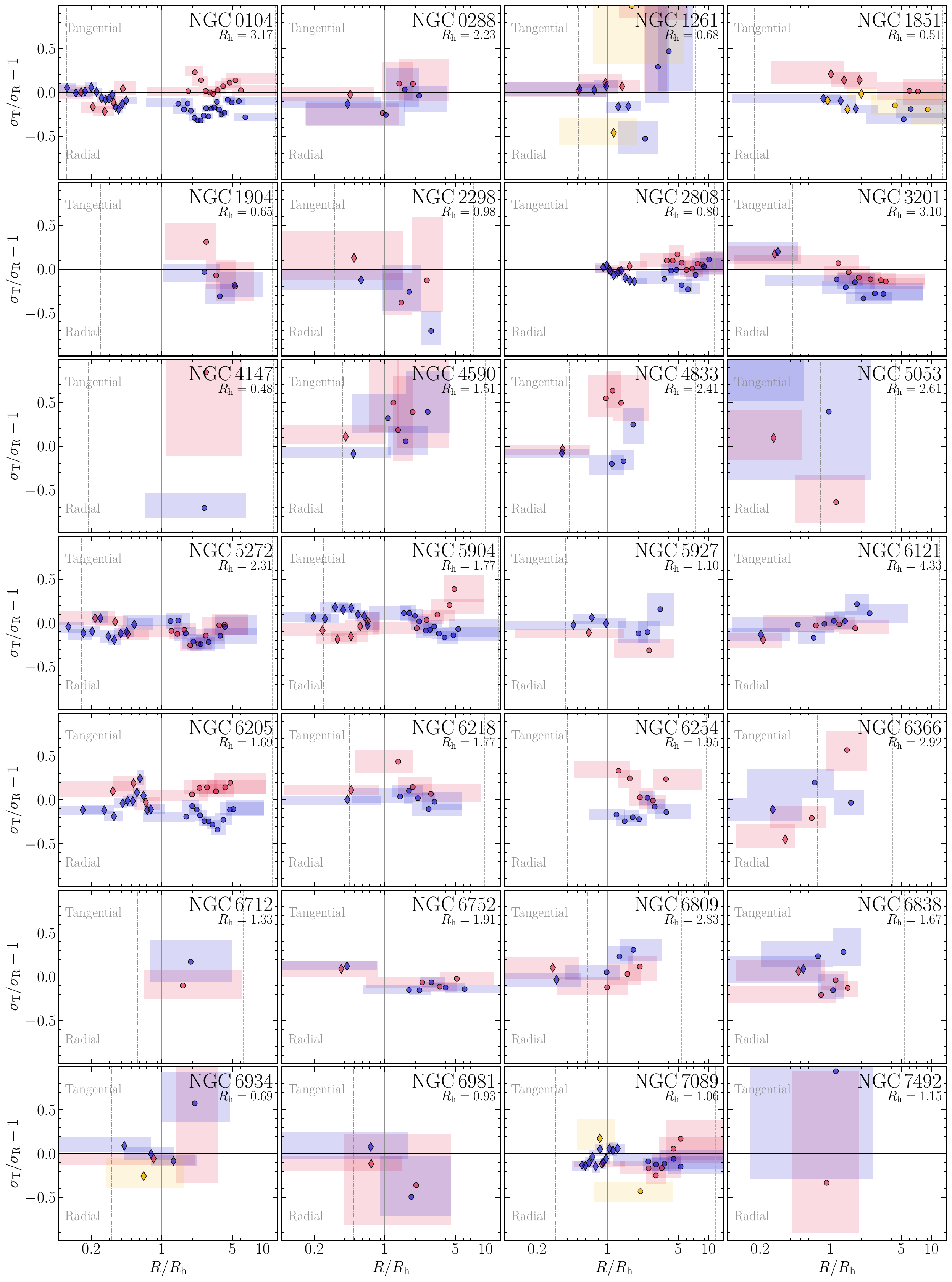}
    \caption{Anisotropy profile, defined as $\beta=\ani$. Red and blue colors indicate results for 1P and 2P stars.}
    \label{fig:beta}
\end{figure*}

We present in this section the individual dynamical profiles of the analyzed clusters. The profiles are displayed in a similar fashion as Fig.~\ref{fig:ind prof}. Mean trends are shown in Fig.~\ref{fig:radmed}-\ref{fig:tanmed}, while dispersion and anisotropy profiles are presented in Fig.~\ref{fig:raddis}-\ref{fig:beta}. Values computed from HST/Gaia data are displayed as crosses and circles respectively, while their uncertainties are indicated by the shaded rectangles. Mean motions and dispersion profiles have been converted to $\mathrm{km/s}$ adopting the clusters' distances derived in \citet{baumgardt2018}. Red and blue colors indicate 1P and 2P stars, while black diamonds represent the values of the 1D-dispersion computed in \citet{libralato2022}. Finally, the locations of the $\rc, \rh, \rt$ radii are indicated by the dash-dotted, solid and dashed lines respectively. $\rt$ is indicated only if contained within the analyzed field of view.

\section{Statistical significance of the observed global profiles.} \label{app:significance}
\begin{figure*}
    \centering
    \includegraphics[width=0.90\textwidth, trim={0cm 2cm 0cm 0cm}, clip]{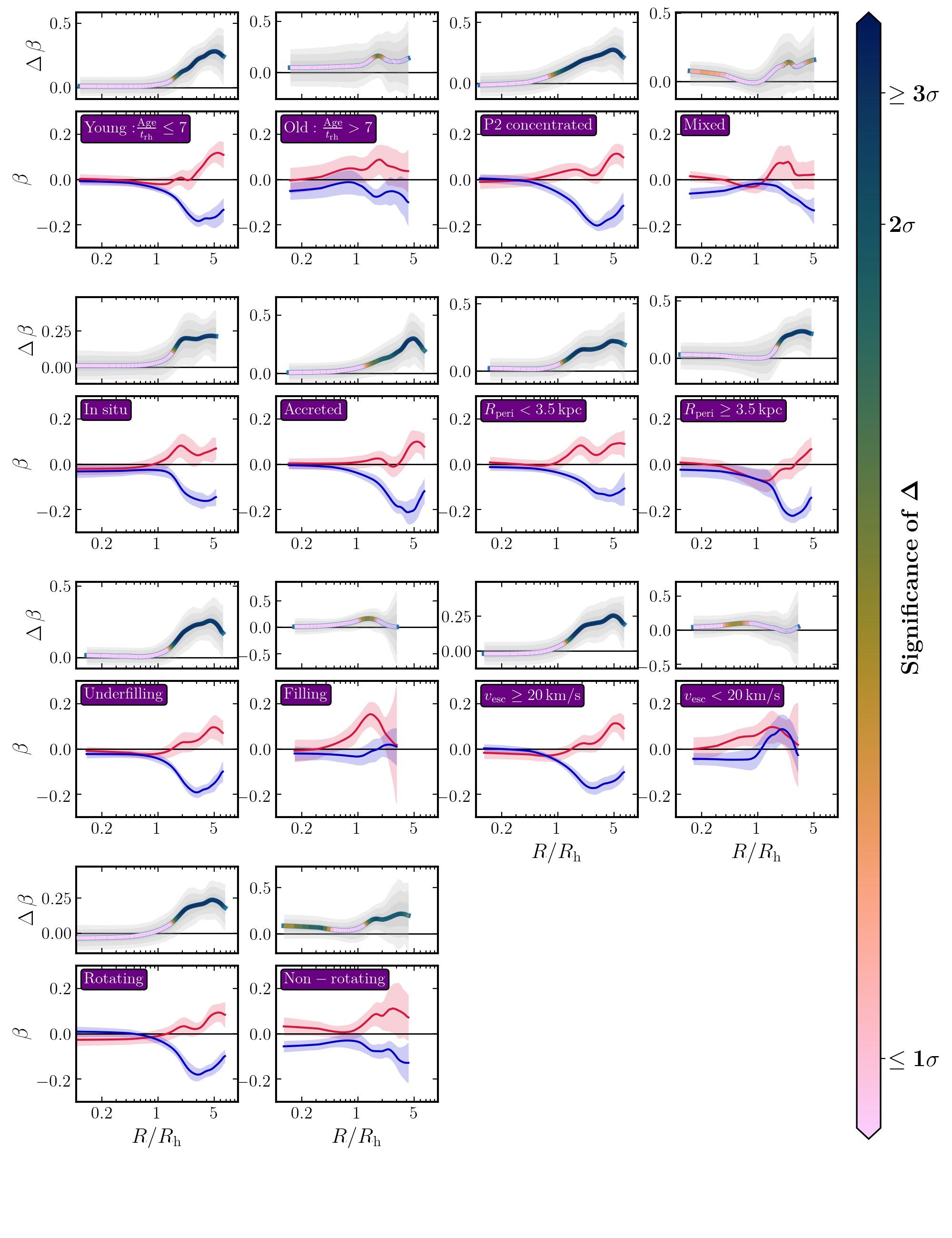}
    \caption{Difference between the 1P and 2P global anisotropy profiles for the groups discussed in Sec.~\ref{sec:discussion} and its statistical significance. In each row, the bottom panels display the global 1P and 2P profiles, along with their uncertainties (red and blue lines and shaded regions). The top panels illustrate the difference between the profiles and its uncertainties. The color of the lines indicates the significance of the difference at each $R/\rh$, as shown by the colorbar. For clarity, points with statistical significance below $1,\sigma$ are represented with the same color.}
    \label{fig:significance1}
\end{figure*}

To assess the statistical significance of the differences in the global dynamical profiles of 1P and 2P shown in Fig.~\ref{fig:stacked all} and ~\ref{fig:comparison}, we adopted the following procedure. We refer for simplicity to the case with all clusters. First we created a 1000 realizations of bootstrapped samples of $y=\sr, \st, \beta$, where each value has bee drawn from a Gaussian with dispersion equal to the observed uncertainty. For each realization we repeated the LOESS fit for 1P and 2P, and computed the difference the between the two fits. Finally, we estimate the $1, 2, 3\sigma$ confidence intervals from the distribution of simulated difference at each $R/\rh$. The three confidence intervals are shown as gray shaded regions in the top panels of Fig.~\ref{fig:significance1}. In addition to this analysis, we also directly quantified the fluctuations introduced solely by the uncertainties. To do this, we repeated the same analysis on two identical distributions, testing the null hypothesis that 1P and 2P stars share the same dynamical profiles, and determined the fraction $(f)$ of simulations with differences larger than the observed ones. Such fraction indicates the probability that uncertainties alone can reproduce a difference between 1P and 2P as large as the observed ones, and thus we computed the significance of the difference as $p=1-f$. The value of $p$ is indicated by the color of the lines in the top panels of Fig.~\ref{fig:significance1}, as indicated by the colorbar.
We stress here that, instead of determining one single value of significance for each 1P/2P dynamical profiles, we compute the statistical significance for each radial coordinate.

\rowcolors{2}{gray!25}{white}
\begin{table*}
    \centering
    \caption{The clusters and groups introduced and analyzed in Sec.~\ref{sec:discussion} are discussed here. The number of 1P, 2P, and Anomalous stars is shown for both HST/Gaia data.}
    \begin{tabular}{ccccccccccc}
        \hline
        \hline
         \textbf{Cluster} & \textbf{Dynamical age} & \textbf{MPs} & \textbf{Rotation} & \textbf{Peri radius}  & $\mathbf{v_\mathrm{esc}}$ & \textbf{Origin} & \textbf{Filling factor} & $\mathbf{N_\mathrm{1P}}$ & $\mathbf{N_\mathrm{2P}}$ & $\mathbf{N_\mathrm{Anomalous}}$\\
        \hline
        \hline
        \textbf{NGC\,0104} & young & p2conc & rot & outer & high & in-situ & underfilling & 314/827 & 1564/1252 & --/-- \\
        \textbf{NGC\,0288} & young & mixed & non rot & inner & low & accreted & filling & 118/146 & 100/135 & --/-- \\
        \textbf{NGC\,1261} & young & mixed & non rot & inner & high & accreted & underfilling & 344/62 & 508/82 & 33/14 \\
        \textbf{NGC\,1851} & young & p2conc & rot & inner & high & accreted & underfilling & 278/85 & 572/103 & 284/71 \\
        \textbf{NGC\,1904} & unknown & p2conc & rot & inner & unknown & accreted & unknown & --/56 & --/128 & --/-- \\
        \textbf{NGC\,2298} & intermediate-old & p2conc & non rot & inner & low & accreted & underfilling & 59/26 & 87/31 & --/-- \\
        \textbf{NGC\,2808} & young & p2conc & rot & inner & high & accreted & underfilling & 449/351 & 1365/397 & --/-- \\
        \textbf{NGC\,3201} & young & p2conc & non rot & outer & low & accreted & underfilling & 66/352 & 96/284 & --/-- \\
        \textbf{NGC\,4147} & unknown & mixed & non rot & inner & unknown & accreted & unknown & --/19 & --/37 & --/-- \\
        \textbf{NGC\,4590} & young & mixed & non rot & outer & low & accreted & underfilling & 106/23 & 219/60 & --/-- \\
        \textbf{NGC\,4833} & young & p2conc & non rot & inner & low & accreted & filling & 156/45 & 225/62 & --/-- \\
        \textbf{NGC\,5053} & young & p1conc & non rot & outer & low & accreted & filling & 32/14 & 23/60 & --/-- \\
        \textbf{NGC\,5904} & young & p2conc & rot & inner & high & accreted & underfilling & 364/306 & 871/318 & --/-- \\
        \textbf{NGC\,5272} & young & p2conc & rot & outer & high & accreted & underfilling & 206/298 & 751/475 & --/-- \\
        \textbf{NGC\,5927} & young & mixed & non rot & outer & high & in-situ & filling & 141/47 & 441/87 & --/-- \\
        \textbf{NGC\,6121} & intermediate-old & mixed & non rot & inner & low & in-situ & filling & 38/104 & 94/240 & --/-- \\
        \textbf{NGC\,6205} & young & mixed & non rot & inner & high & accreted & underfilling & 221/230 & 1009/379 & --/-- \\
        \textbf{NGC\,6218} & intermediate-old & mixed & rot & inner & low & in-situ & filling & 131/126 & 179/202 & --/-- \\
        \textbf{NGC\,6254} & intermediate-old & p2conc & non rot & inner & high & in-situ & filling & --/218 & --/255 & --/-- \\
        \textbf{NGC\,6366} & intermediate-old & mixed & non rot & inner & low & in-situ & filling & 29/55 & 43/54 & --/-- \\
        \textbf{NGC\,6712} & unknown & p1conc & non rot & inner & unknown & in-situ & unknown & --/49 & --/42 & --/-- \\
        \textbf{NGC\,6752} & intermediate-old & mixed & rot & inner & high & in-situ & underfilling & 103/178 & 265/360 & --/-- \\
        \textbf{NGC\,6809} & young & mixed & norot & inner & low & in-situ & filling & 47/60 & 117/118 & --/-- \\
        \textbf{NGC\,6838} & intermediate-old & mixed & non rot & outer & low & in-situ & filling & 67/78 & 32/50 & --/-- \\
        \textbf{NGC\,6934} & young & mixed & norot & inner & high & in-situ & underfilling & 205/25 & 364/89 & 30/-- \\
        \textbf{NGC\,6981} & intermediate-old & p2conc & non rot & inner & low & accreted & underfilling & 201/53 & 186/50 & --/-- \\
        \textbf{NGC\,7089} & young & mixed & rot & inner & high & accreted & underfilling & 273/165 & 941/329 & 57/20 \\ 
        \textbf{NGC\,7492} & unknown & mixed & non rot & inner & unknown & accreted & unknown & --/27 & --/49 & --/-- \\
        \hline
        \hline
    \end{tabular}
\label{tab:tab1}
\end{table*}

\bsp	
\label{lastpage}
\end{document}